\documentclass{article}

\usepackage{PRIMEarxiv}

\usepackage[ruled]{algorithm2e} 

\usepackage{amsfonts}       
\usepackage{amsmath}
\usepackage{amssymb}
\usepackage{amsthm}
\usepackage{authblk}
\usepackage{booktabs}       
\usepackage{boxedminipage}
\usepackage{caption} 
\usepackage{cleveref}
\usepackage{comment}
\usepackage{enumitem}
\usepackage{fancyhdr}       
\usepackage{graphicx}       
\usepackage[utf8]{inputenc} 
\usepackage{lipsum}
\usepackage{microtype}      
\usepackage{multirow}
\usepackage{natbib}  
\usepackage{nicefrac}       
\usepackage{pifont}
\usepackage{pgfplots}
\usepackage[T1]{fontenc}    
\usepackage{thm-restate}
\usepackage{threeparttable}
\usepackage{tikz}
\usepackage{url}            
\usepackage{xcolor}         
\usepackage{xr}

\setlist{nolistsep}
\usetikzlibrary{arrows}
\usetikzlibrary{patterns}
\SetKw{Break}{break}

\newtheorem{definition}{Definition}%
\newtheorem{theorem}{Theorem}%
\newtheorem{proposition}{Proposition}%
\newtheorem{example}{Example}

\newcommand{\cmark}{\ding{51}} 
\newcommand{\xmark}{\ding{55}} 

\allowdisplaybreaks

\title{\bf A Fair and Optimal Approach to Sequential Healthcare Rationing}

\author[1,2]{Zhaohong Sun}

\affil[1]{Kyushu University, Japan}
\affil[2]{CyberAgent, Japan}

\date{\vspace{-10mm}}

\begin{document}

\maketitle

\begin{abstract}
The COVID-19 pandemic underscored the urgent need for fair and effective allocation of scarce resources, from hospital beds to vaccine distribution. In this paper, we study a healthcare rationing problem where identical units of a resource are divided into different categories, and agents are assigned based on priority rankings.
We first introduce a simple and efficient algorithm that satisfies four fundamental axioms critical to practical  applications: eligible compliance, non-wastefulness, respect for priorities, and maximum cardinality. This new algorithm is not only conceptually simpler but also computationally faster than the Reverse Rejecting rules proposed in recent work.
We then extend our analysis to a more general sequential setting, where categories can be processed both sequentially and simultaneously. For this broader framework, we introduce a novel algorithm that preserves the four fundamental axioms while achieving additional desirable properties that existing rules fail to satisfy. Furthermore, we prove that when a strict precedence order over categories is imposed, this rule is the unique mechanism that satisfies these properties.
\end{abstract}

\section{Introduction}
The COVID-19 pandemic underscored the urgent need for fair and efficient healthcare rationing as hospitals around the world grappled with severe shortages of critical resources such as ventilators, ICU beds, and vaccines, especially at the onset of the outbreak. In these crises, decision-makers were confronted with the challenge of balancing two essential considerations: fairness, which ensures that resources are allocated according to transparent and justifiable criteria, and optimality, which seeks to distribute resources in a way that maximizes social benefit.

In an influential paper, \citet{PSUY24a} develops a comprehensive theory of reserve system design, offering significant implications for both policy and practical healthcare rationing. Prior to this work, healthcare rationing guidelines predominantly advocated for a priority system, where patients are ranked in a single priority order and resources are allocated based on this list. However, relying solely on a single priority list to allocate scarce resources has contributed to the shortcomings of existing pandemic rationing protocols. To address this, the authors introduce the concept of a \emph{reserve system}, where identical units of resources are allocated across multiple categories, each with its own specific priority order for resource distribution within that category.

One key aspect of the reserve system is that an agent may be eligible for multiple categories. For instance, healthcare professionals might qualify for a vaccine both through the essential-worker category and the unreserved category open to all patients. In such cases, it is essential to specify which category will be used for allocation. \citet{PSUY24a} propose utilizing the classical Deferred Acceptance (DA) algorithm, assuming that agents have preferences over categories. They demonstrate that the DA algorithm satisfies three critical axioms, inspired by real-world applications: \emph{eligibility compliance} (ensuring that resources designated for a particular category are allocated only to eligible individuals), \emph{non-wastefulness} (ensuring that no resource remains unallocated while there are individuals who need it), and \emph{respect for priorities} (ensuring that resources are allocated according to category-specific priorities, in alignment with ethical principles and guidelines).

\citet{AzBr24a} builds upon the work of \citet{PSUY24a} by introducing the Reverse Rejecting (REV) rule, which satisfies the three key axioms mentioned earlier, along with another fundamental axiom of \emph{maximum cardinality}, ensuring that the greatest number of individuals receive the allocated resources. However, the REV rule may have several limitations. First, it heavily depends on a baseline priority order over individuals, and variations in this order can lead to significant differences in the resulting allocation, without a clear method for determining the most appropriate baseline. Second, it is challenging to generalize the rule to sequential reserve systems, which are central to practical applications but have not been fully explored yet. Third, it can be computationally expensive, as it requires calculating a maximum matching $|I|$ times (with $I$ representing the set of agents), which may not be feasible for large-scale markets.

In this paper, we revisit the reserve system proposed by \citet{PSUY24a}, and our first research question is: \emph{How can we design a computationally efficient algorithm that satisfies these four fundamental axioms without relying on an additional baseline order?}

In many real-life applications of reserve systems, institutions allocate units sequentially by processing reserve categories one at a time according to their category-specific priority order. \citet{PSUY24a} formalize this framework as a sequential reserve setting, where a precedence order over categories determines the processing sequence. However, they do not provide an optimal solution for maximizing the number of matched agents which achieves all the fundamental axioms. Instead, they show that when all categories share a common baseline ranking over agents, maximum cardinality can be achieved using the existing Smart Reserve Rule \citep{SoYe22a}. Yet, enforcing a uniform baseline ordering restricts the system to homogeneous priorities, which may not be suitable for complex, heterogeneous settings. 
\citep{AzBr24a} examine a specific form of sequential reserve allocation where open categories (eligible to all agents) are processed either before or after reserved categories. They propose a method that combines the Reverse Rejecting Rule with the Smart Reserve Rule to achieve maximum matching in this setting. However, their approach does not generalize to the broader sequential reserve framework.  

This raises a fundamental yet unresolved research question—one of the central issues explored in this paper: \emph{Can we design an algorithm that satisfies the four fundamental axioms in a general reserve system where categories are processed both sequentially and simultaneously?}

\begin{table}[tb]
\centering
\resizebox{\columnwidth}{!}{
\begin{tabular}{ccccccc}
\toprule
\textbf{class} & \textbf{axioms / properties} & \textbf{DA} & \textbf{REV} & \textbf{SREV} & \textbf{MMA} & \textbf{SCU} \\
\midrule
\multirow{4}{*}{fundamental} 
        & eligibility compliance  & \cmark & \cmark  &  \cmark & \cmark & \cmark \\
        & non-wastefulness  & \cmark & \cmark  &  \cmark & \cmark & \cmark \\
        & respect for priorities (over agents)   & \cmark & \cmark  &  \cmark & \cmark & \cmark \\
        & maximum cardinality  & \xmark & \cmark  &  \cmark & \cmark & \cmark \\
 \midrule
\multirow{3}{*}{consistency} 
         & independence of baseline   & \cmark & \xmark  & \xmark & \cmark & \cmark \\
         & consistent matched agents & \cmark$^1$ & \cmark  & \cmark & \xmark & \cmark \\
         & consistent matching & \cmark$^1$ & \xmark  & \xmark & \xmark & \cmark \\
 \midrule
 \multirow{2}{*}{incentive}  
        & no incentive to hide & \cmark &  \cmark & \cmark & -- & \cmark \\
        & respect for improvements & \cmark &  \cmark & \cmark & -- & \cmark \\
\midrule
 \multirow{3}{*}{sequential} 
        & maximum beneficiary assignment  & \xmark &  \xmark & \xmark$^2$ & \xmark & \cmark \\
        & order preservation by swap  & \xmark &  \xmark & \xmark$^2$ & \xmark & \cmark \\
        & respect for precedence (over categories) & \xmark & \xmark & \xmark & \xmark & \cmark \\
\bottomrule
\end{tabular}
}
\caption{Comparison of five algorithms including Deferred Acceptance (DA) \citep{PSUY24a}, Reverse Rejecting (REV) \citep{AzBr24a}, Smart Reverse Rejecting (SREV) \citep{AzBr24a}, Maximum Matching Adjustment (MMA), and Sequential Category Update (SCU).
$^1$ For DA, we assume that ties in agents' preferences are broken in a predefined manner, such as based on the indices of the categories. 
$^2$ For SREV, these two properties can be satisfied under a restrictive setting where an open category is processed in the beginning or in the end.}
\label{tab:comparison:sequential}
\end{table}

We summarize our contributions as follows, with Table~\ref{tab:comparison:sequential} comparing our newly proposed algorithms with existing solutions across four classes of axioms and properties.  

\begin{itemize}
    \item \textbf{First,} we introduce the Maximum Matching Adjustment (MMA) rule, which not only satisfies the four fundamental axioms (Theorem~\ref{theo:MMA:properties}) but also outperforms the only known algorithm with these properties in terms of both simplicity and computational efficiency (Theorem~\ref{theo:MMA:time}). Furthermore, we characterize these axioms by proving that a matching satisfies all four conditions if and only if it is induced by the MMA rule (Theorem~\ref{theo:characterization:MMA}).
    
    \item \textbf{Second,} we present the Sequential Category Updating (SCU) rule, which is capable of handling both the basic reserve system and the sequential reserve system. In addition to satisfying the four fundamental axioms and the three axioms specific to the sequential model, it achieves three consistency properties and two incentive properties (Theorem~\ref{theorem:SCU:properties}). Moreover, under a strict precedence order over categories, the SCU rule is the unique mechanism that satisfies eligibility compliance, maximum cardinality, non-wastefulness, respect for priorities, and respect for precedence (Theorem~\ref{theorem:SCU:unique}).

    \item \textbf{Third,} while previous work has primarily relied on bipartite graphs, we demonstrate that implementing the rule using flow networks offers a more intuitive approach. This method can significantly improve computational efficiency when grouping agents based on their eligible categories. For completeness, we also provide a fast implementation of the SCU rule using bipartite graphs.
\end{itemize}


\section{Related Work}
\label{section:literature}

The design of reserve systems for health rationing was first introduced by \citep{PSUY21a, PSUY24a}, where the authors developed a general theoretical framework with significant policy implications, particularly during and after the COVID-19 pandemic. They demonstrated that the classical deferred acceptance algorithm satisfies three fundamental axioms: eligibility compliance, non-wastefulness, and respect for priorities. Building on this framework, \citep{AzBr21a, AzBr24a} introduced the first known algorithm that satisfies the fourth fundamental axiom: maximal cardinality. While their work laid the foundation for sequential reserve systems in health rationing, they did not address whether it is possible to design an algorithm that satisfies all four fundamental axioms when categories are processed sequentially. In this paper, we provide an affirmative answer to this question and introduce the Sequential Category Updating (SCU) rule, which not only achieves the four fundamental axioms but also nice properties in sequential reserve systems.

The precedence order over categories has been studied in various matching settings.  
\citet{KoSo16a} investigate a matching problem with slot-specific priorities and introduce the concept of a precedence order for slots, which corresponds to categories in reserve matching. Their model is more general, as it considers agents with unit demand who are matched to branches with multiple slots, each with its own priority ranking over agents. A branch then selects agents by filling its slots sequentially according to a predetermined precedence order.  
\citet{DKPS18a} examine the role of precedence order in admissions reserves, where different types of seats are allocated sequentially based on a specified precedence order. They show that either lowering the precedence of reserve seats at a school or increasing the school's reserve size weakly increases the number of assignments to the reserve group at that school. 
\citet{PRS23a} explore public misconceptions about reserve systems in affirmative action policies, highlighting how a lack of awareness about processing order leads individuals to conflate policies with different levels of affirmative action.  
However, achieving maximum cardinality is not the primary objective in these works.

The reserve system has been studied in other contexts, and the healthcare rationing problem shares similarities with the literature on school choice with diversity goals. In the latter setting, schools reserve seats for different categories, such as gender, socioeconomic status, or neighborhood affiliation. \citet{HYY13a, EHYY14a} studied reserve systems in school choice settings, where each student belongs to a single category and a baseline priority is applied across all categories. Subsequent work expanded this framework to more general settings where agents may belong to multiple categories \citep{ BCC+19a, CEE+19a, KHIY17a, GNKR19a, Aziz19b, AGS20a}. However, a key difference lies in the nature of preferences: in healthcare rationing, agents have dichotomous preferences, whereas in school choice, students have strict preferences over schools.

More generally, this paper falls within the broader research area of matching with distribution constraints. Various forms of distributional constraints have been explored, including matching with minimum quotas \citep{BFIM10a, HHKS+17a}, matching with regional quotas \citep{KaKo15a, GKH+15a, KaKo17a, GIKY+16a}, and matching under complex constraints \citep{KTY18a,KaKo23a}.

\section{Preliminaries}
\label{sec:Preliminaries}

We adopt the model from \citet{PSUY21a,PSUY24a} and \citet{AzBr21a,AzBr24a}. A reserve system $R = (I, C, q, \succ)$ consists of a set of agents $I$ and a set of reserve categories $C$. There are multiple identical, indivisible resource units to allocate, with each agent requiring exactly one unit.

Each category $c \in C$ has $q_c$ reserved units, where the reserved capacity vector is denoted as $q = (q_c)_{c \in C}$. Every category $c$ has a strict priority ranking $\succ_c$ over the agents in $I \cup \{\emptyset\}$. An agent $i$ is \emph{eligible} for category $c$ if $i \succ_c \emptyset$. The overall priority profile across all categories is denoted by $\succ$.

A matching $\mu: I \cup C \to I \cup C \cup \{\emptyset\}$ is a function that assigns each agent to a category or leaves them unmatched, denoted by $\emptyset$, while respecting capacity constraints. For each category $c \in C$, the number of agents assigned to $c$, denoted by $|\mu(c)|$, must not exceed its reserved capacity $q_c$. For an agent $i \in I$, $\mu(i) = c$ indicates that $i$ is assigned a unit of resource reserved for category $c$, while $\mu(i) = \emptyset$ means that $i$ remains unmatched.


Given a reserve system $R = (I, C, \succ, q)$, a corresponding eligibility graph $G = (I \cup C, E)$ is a bipartite graph with vertices consisting of agents $I$ and reserved categories $C$. An edge exists between an agent $i$ and a reserved category $c$ if the agent $i$ is eligible for category $c$. 
Thus the set of edges is denoted as $E = \{(i, c): i \succ_c \emptyset\}$. 
For each node $c \in C$, there are $q_c$ reserved units. 

\begin{example}
\label{example:instance}
Consider a reserve system with three agents $I = \{i_1, i_2, i_3\}$ and two categories $C = \{c_1, c_2\}$. Each category has a capacity of 1, and the priority profile is given in the Figure~\ref{fig:example}.
\begin{figure}[htb]
\begin{center}
\begin{tikzpicture}[-,>=stealth',shorten >=1pt,auto,node distance=1cm,
thick,main node/.style={circle,fill=white!15,draw,minimum size=0.5cm,inner sep=0pt}, scale=0.8]

\node[main node] at (0, 2) (i1)  {$i_1$};
\node[main node] at (0, 1) (i2)  {$i_2$};
\node[main node] at (0, 0) (i3)  {$i_3$};

\node[main node] at (3, 1.5) (c1)  {$c_1$};
\node[main node] at (3, 0.5) (c2)  {$c_2$};

\node[align=left] at (7.5, 1.5) (succ1)  {$\succ_{c_1}: i_2 \succ i_1 \succ \emptyset \succ i_3$};
\node[align=left] at (7.5, 0.5) (succ2)  {$\succ_{c_2}: i_2 \succ i_3 \succ \emptyset \succ i_1$};

\draw[] (i1) -- (c1);
\draw[] (i2) -- (c1);
\draw[] (i2) -- (c2);
\draw[] (i3) -- (c2);

\end{tikzpicture}
\end{center}
\caption{The eligibility graph for Example~\ref{example:instance}, with the priority profile shown on the right.}
\label{fig:example}
\end{figure}
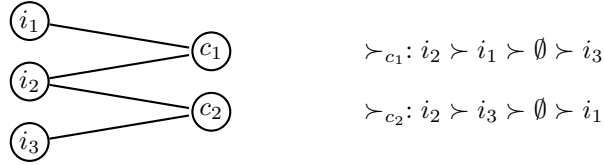
\end{example}

\subsection{Fundamental Axioms}  
We now introduce four fundamental axioms that a desirable matching should satisfy, as proposed and formalized in previous work. These axioms serve as essential principles for designing reserve systems in health rationing. For a more detailed discussion on the real-life implications of these axioms, we refer readers to \citep{AzBr24a,PSUY24a}.

The first axiom, \emph{eligibility compliance}, asserts that resources allocated to any category should only be assigned to eligible individuals. For example, in vaccine distribution, vaccines reserved for elderly citizens should not be administered to young adults; similarly, in school choice, seats reserved for minorities should not be assigned to the majority.

We use $\Psi$ to denote the set of all matchings that satisfy the requirement of eligibility compliance. For the remainder of the paper, we will focus solely on matchings in $\Psi$ and may occasionally omit the explicit mention of $\Psi$.

\begin{definition}[Eligibility Compliance]
\label{def:eligibility}
A matching $\mu$ complies with eligibility requirements if, for any category $c \in C$ and any agent $i \in I$, if $\mu(i) = c$, then $i \succ_c \emptyset$.
\end{definition}

The second axiom, \emph{respect for priorities}, expresses the principle that within each category, resources should be allocated according to the priority order of individuals. This property is also known as \emph{elimination of justified envy} in the literature on school choice \citep{AbSo03b,CEE+22a,STY+23a}, or \emph{fairness} in the literature on two-sided matching markets \citep{GIKY+16a,AzSu21a,KaKo23a}.

\begin{definition}[Respect for Priorities]
    \label{def:respect_priorities}
    A matching $\mu$ respects priorities if, for any agents $i, i' \in I$ and any category $c \in C$, if $\mu(i) = c$ and $\mu(i') = \emptyset$, then $i \succ_c i'$.
\end{definition}

The third axiom, \emph{non-wastefulness}, is a mild requirement of efficiency. It 
asserts that no resource unit should remain unallocated as long as there is an eligible individual who desires it. In other words, every eligible agent should be matched to a category with available resources if such a matching is possible.

\begin{definition}[Non-wastefulness]
    \label{def:NW}
    A matching $\mu$ satisfies non-wastefulness if, for any agent $i \in I$ and any category $c \in C$, if $i \succ_c \emptyset$ and $\mu(i) = \emptyset$, then $|\mu(c)| = q_c$.
\end{definition}

The fourth axiom, \emph{maximum cardinality}, stipulates that the number of matched agents should be maximized while ensuring eligibility compliance. This axiom was mentioned by \citet{PSUY21a} and formalized by \citet{AzBr21a}. 
A non-wasteful matching ensures that no eligible agent is left unmatched, but it does not necessarily guarantee the largest possible number of matched agents. On the other hand, a maximum-cardinality matching implies non-wastefulness.

\begin{definition}[Maximum Cardinality]
    \label{def:MS}
    A matching $\mu$ is a \emph{maximum-cardinality} matching if matches the largest number of agents among all matchings that satisfy the eligibility requirements, i.e., $\mu \in \arg\max_{\mu' \in \Psi} |\mu'|$.
\end{definition}

\subsection{Additional Properties for Algorithms}

An algorithm $\Gamma: \mathcal{R} \rightarrow \Pi$ maps each reserve system $R\in \mathcal{R}$ to a matching $M \in \Pi_R$, where $\mathcal{R}$ is the set of all reserve systems, and $\Pi_R$ is the set of possible matchings for a given instance $R$.

We say that an algorithm $\Gamma$ satisfies an axiom if, for every instance $R \in \mathcal{R}$, the resulting matching $\Gamma(R)$ adheres to that axiom. Given an agent $i$, we denote by $\Gamma(R, i)$ the assignment of agent $i$ under the matching produced by $\Gamma$.

Next, we introduce additional properties that an algorithm should satisfy, focusing primarily on the incentives for agents to misreport their eligible categories and the consistency of the algorithm's output given the same instance.

Recall that agents perceive all resource units as identical and do not have preferences over the categories to which they are assigned. An algorithm satisfies the no-incentive-to-hide property if no agent can benefit by hiding their eligibility for a category \citep{AyBo20a,AzSu21a}. In other words, there is no agent who is unmatched when reporting all of their eligible categories but becomes matched when reporting only a subset of those categories.

\begin{definition}[No Incentive to Hide]
    \label{def:hide}
An algorithm $\Gamma$ does not incentivize agents to hide their eligible categories if, for a reserve system $R$ and an agent $i$, whenever $\Gamma(R, i) = \emptyset$, it must also hold that $\Gamma(R', i) = \emptyset$ for any modified reserve system $R'$ in which agent $i$ hides some of their eligible categories, while all other elements in $R'$ and $R$ are unchanged.
\end{definition}

\citet{BaSo99a} studied a stronger property than the no-incentive-to-hide property, called \emph{respect for improvements}. This property asserts that an agent is never worse off under an algorithm when their priority increases in one or more categories. More precisely, if an agent is assigned a unit under a given priority profile, they should still be assigned a unit if their priority in each category increases or remains the same, while the priorities of all other agents remain unchanged.

\begin{definition}[Respect for Improvements]
\label{def:axiom:improvement}
An algorithm $\Gamma$ \textit{respects improvements} if, for any agent $i$, 
whenever $\Gamma(R, i) \neq \emptyset$, it must also hold that $\Gamma(R', i) \neq \emptyset$ for any modified reserve system $R'$ in which agent $i$'s priority is weakly higher in at least one category, while the priorities of all other agents remain unchanged.
\end{definition}

\begin{proposition}
    \label{proposition:connection:manipulation}
    Respecting improvements implies no incentive to hide.
\end{proposition}

\begin{proof}
Suppose an algorithm $\Gamma$ satisfies the respect for improvements property. The contrapositive of this definition states: if $\Gamma(R_1, i) = \emptyset$, then $\Gamma(R_2, i) = \emptyset$, where $R_1$ is a modified reserve system derived from $R_2$ in which agent $i$'s priority is weakly higher in at least one category, while the priorities of all other agents remain unchanged.

Notice that when agent $i$ hides an eligible category $c$, it is equivalent to lowering their priority for category $c$ below $\emptyset$ (i.e., treating $i$ as ineligible for $c$). Therefore, if $\Gamma(R_1, i) = \emptyset$, it follows that $\Gamma(R_2, i) = \emptyset$ in the modified reserve system $R_1$, where agent $i$'s priority has been weakly increased in at least one category. This condition satisfies the definition of no-incentive-to-hide.
\end{proof}

In previous work \citep{AzBr24a}, a baseline priority order is imposed on all agents. Based on this order, the proposed algorithm determines whether each agent can be matched, ensuring compliance with four fundamental axioms. As discussed in the Introduction, this approach has several limitations, for example, altering the baseline priority can lead to significantly different outcomes, where agents matched under one priority may become unmatched under another. 

The following property requires that, for the same reserve system, if an agent is matched under the algorithm, they should remain matched regardless of the baseline priority used.

\begin{definition}[Independence of Baseline Priority] \label{def:independence} 
An algorithm $\Gamma$ is said to be independent of the baseline priority if, for a given reserve system $R$ and any two baseline priorities $\pi$ and $\pi'$, it holds that for all $i \in I$, whenever $\Gamma(R, i, \pi) \neq \emptyset$, then $\Gamma(R, i, \pi') \neq \emptyset$. \end{definition}

The following property ensures consistency in the algorithm with respect to the matched agents: for the same input instance, the set of matched agents produced by the algorithm remains unchanged, even if they may be assigned to different categories.

\begin{definition}[Consistent Matched Agents] \label{def:consistent_agents} An algorithm $\Gamma$ is said to be consistent with respect to matched agents if, for two given reserve systems $R$ and $R'$, where $R = R'$, for all $i \in I$, if $\Gamma(R, i) \neq \emptyset$, then $\Gamma(R', i) \neq \emptyset$.
\end{definition}

As argued by \citep{PSUY24a}, it is not always sufficient to specify merely who receives a unit; it is equally important to specify through which category agents receive their units. For example, it is crucial when we need to specify the minimum priority required for an agent to be assigned a unit of resource in each category. 
The next property is stricter than the Consistent Matched Agents property, ensuring that each agent is matched to the same category for the same input instance.

\begin{definition}[Consistent Matching] \label{def:consistent_matching} 
An algorithm $\Gamma$ is said to be consistent with respect to outcomes if, for two given reserve systems $R$ and $R'$ with $R = R'$, it holds that $\Gamma(R) = \Gamma(R')$. 
\end{definition}

\begin{proposition}
    \label{proposition:connection:consistency}
    The Consistent Matching property implies the Consistent Matched Agents property, but not vice versa. 
\end{proposition}

\subsection{Previous Algorithms}
We now provide a brief description of previous algorithms.

\citet{PSUY24a} consider the Individual Proposing Deferred Acceptance (DA) algorithm, which involves artificially generating a strict preference order for each agent over all categories, and then applying the classical deferred acceptance algorithm to the induced instance. The formal description of the DA algorithm is provided in Algorithm~\ref{alg:DA}. 

\begin{algorithm}[tb]
    \caption{Deferred Acceptance (DA) Algorithm}
    \label{alg:DA}
    \KwIn{A reserve system $R = (I, C, \succ, q)$}
    \KwOut{A matching $\mu$}
    
    Each agent $i \in I$ submits a strict preference ranking $\succ_i$ over eligible categories \;

    \While{some agents remain unmatched and can still propose}{
        \textbf{Agents Propose:} Each unmatched agent $i \in I$ applies to her most preferred category among those for which she is eligible and has not yet been rejected, if such a category exists \;
        \textbf{Categories Select:} Each category $c \in C$ tentatively selects agents from the new applicants in the current round and those previously matched in the last round, prioritizing them according to $\succ_c$ until its capacity $q_c$ is reached. Any unselected applicants are rejected \; 
    }
    \Return the final assignment $\mu$.
\end{algorithm}

\begin{proposition}[\citep{PSUY24a}]
    \label{theorem:DA:properties}
The Deferred Acceptance Algorithm returns a matching that satisfies eligibility compliance, non-wastefulness, and respect for priorities.
\end{proposition}

\citet{AzBr24a} proposed the Reverse Rejecting (REV) Rule, which operates based on a baseline ordering $\pi$ over all agents and processes them iteratively in reverse order of $\pi$. An agent $i$ is rejected only if a matching can still be found without $i$, while satisfying all four fundamental properties. Specifically, $i$ is removed if there exists a maximum cardinality matching in an updated eligible graph where: (1) agent $i$ is excluded, and (2) all edges $(j, c)$ where $i \succ_c j$ are also removed. In other words, an agent can be safely rejected only if maximal cardinality can still be achieved without them, and removing the corresponding edges does not violate the priority respect requirement. The formal description of the DA algorithm is provided in Algorithm~\ref{alg:REV}.

\begin{proposition}[\citep{AzBr24a}]
    \label{theorem:REV:properties}
The Reverse Rejecting (REV) Rule returns a matching that satisfies eligibility compliance, non-wastefulness, respect for priorities, and maximum cardinality.
\end{proposition}

\begin{algorithm}[tb]
    \caption{Reverse Rejecting (REV) Rule}
    \label{alg:REV}
    \KwIn{A reserve system $R = (I, C, \succ, q)$ and a baseline order $\pi$ over the agents $I$}
    \KwOut{A matching $\mu$}

 Construct an initial eligible graph $G$ between $I$ and $C$. Compute a maximum cardinality matching and denote its size as $m$\;
 
    Initialize the set of rejected agents: $R \leftarrow \emptyset$\;

    \ForEach{agent $i$ in \textbf{reverse} order of $\pi$}{
        Construct the modified eligible graph $G^i$ by removing agent $i$ and all edges $(j, c)$ where $i \succ_c j$\;
        
        \If{$G^i$ admits a maximum matching of size $m$}{
            Add $i$ to the rejected agents: $R \leftarrow R \cup \{i\}$\;
            Update $G$ by removing agent $i$ and all edges $(j, c)$ where $i \succ_c j$ for any $c \in C$\;
        }
    }

    Compute a maximum cardinality matching $\mu$ in the final graph $G$\;
\end{algorithm}

\section{A Simple Algorithm That Satisfies Four Fundamental Axioms}
\label{sec:MMA}

In this section, we address the first research  question: how to design an efficient algorithm that achieves the four fundamental axioms,  without additionally relying on a baseline order.
We next introduce such an algorithm that not only satisfies these axioms but is also simpler and significantly faster than the Reverse Rejecting (REV) Rule \citep{AzBr24a}.

The conception behind the newly proposed rule, Maximum Matching Adjustment (MMA), is straightforward. The MMA rule consists of two steps, with its formal description provided in Algorithm~\ref{alg:MMA}. It begins with any maximum size matching and progressively eliminates pairs of agents that violate the priority-respecting requirement, i.e., by allowing an unmatched agent to displace a matched agent with lower priority.


In the first step, it constructs an eligibility graph $G$ and computes a maximum-size matching, denoted by $\mu$. Various algorithms can be employed to compute a maximum matching for a bipartite graph, including the well-known Hopcroft–Karp algorithm~\citep{HoKa73a}.

In the second step, the matching $\mu$ is gradually adjusted to satisfy the axiom of respecting priorities. Specifically, for each unmatched agent $i$, the algorithm checks each eligible category $c$ for agent $i$ to determine whether $i$ has a higher priority than the agent $i'$, who is matched to category $c$ in $\mu$ and has the lowest priority among the agents matched to category $c$ in $\mu$. If $i$ has a higher priority than $i'$, then $i$ replaces $i'$ in category $c$, and $i'$ becomes unmatched, and the matching $\mu$ is updated accordingly. Then the algorithm moves on to the next unmatched agent. This process is repeated until all unmatched agents and their eligible categories have been checked. 

\begin{algorithm}[tb]
\caption{Maximum Matching Adjustment (MMA) Rule}
\label{alg:MMA}
\KwIn{A reserve system $R = (I, C, \succ, q)$}
\KwOut{A matching $\mu$}
Construct the eligibility graph $G = (I \cup C, E)$ for the reserve system $R$\;
Find a maximum-size matching in $G$, denoted as $\mu$\;

\While{there exists an unmatched agent $i$}{
    \While{agent $i$ is eligible for some category $c$ that has not yet been considered}{
        Let $i'$ be the agent with the lowest priority currently matched to $c$ in $\mu$\;
        \If{agent $i$ has higher priority than $i'$}{
            Update $\mu$ by setting $\mu(i) \gets c$ and  $\mu(i') \gets \emptyset$ \;
            \textbf{break}
        }
    }
}
\Return $\mu$\;
\end{algorithm}

\begin{example}
    \label{example:MMA}
We now illustrate how the MMA rule operates on Example~\ref{example:instance}. Suppose the initial maximum matching $\mu$ is given by $\mu(i_1) = c_1$ and $\mu(i_3) = c_2$. The unmatched agent $i_2$ then proposes to $c_1$ and gets matched, causing $i_1$ to become unmatched. Since $i_1$ has no remaining eligible categories to propose to, MMA terminates.  
Note that if $i_2$ had proposed to $c_2$ instead of $c_1$, MMA would have produced a different matching where $i_2$ is matched to $c_2$, displacing $i_3$.
\end{example}

\begin{restatable}{theorem}{TheoremMMA}\label{theo:MMA:properties}
The MMA algorithm returns a matching that achieves eligibility compliance, respect for priorities, non-wastefulness, and maximum cardinality.
\end{restatable}


\begin{proof}
We now demonstrate that the matching $\mu$ returned by the MMA algorithm satisfies the four fundamental properties.
\\
\textbf{Eligibility Compliance:}  
The algorithm starts by constructing a compact eligibility graph that links agents to the reserved units they are eligible for. Throughout the process, agents are only matched with units from categories for which they qualify, thereby ensuring eligibility compliance.
\\
\textbf{Respect for Priorities:}  
Assume, for the sake of contradiction, that the outcome $\mu$ produced by the MMA algorithm does not respect priorities. This implies that there exist two agents $i, i' \in I$ and a category $c \in C$ such that $\mu(i) = \emptyset$, $\mu(i') = c$, and $i \succ_c i'$.  Two cases arise:  
\begin{enumerate}
    \item Agent $i$ was unmatched at the end of Step 1: \\
    In this case, during Step 2, agent $i$ must have considered category $c$. At that time, the lowest-priority agent matched to $c$, denoted by $i^*$, must have had a higher priority than $i$. Otherwise, $i$ would have displaced $i^*$ and been matched to $c$.
    \item Agent $i$ was initially matched to category $c$ at the end of Step 1: \\
    Here, during Step 2, agent $i$ must have been displaced by another agent $i^*$ with a higher priority for category $c$. This displacement ensures that $i$ could not remain matched to $c$.
\end{enumerate}

Throughout the execution of the MMA algorithm, no category $c$ is ever matched to an agent with a lower priority than any agent who has been rejected. Consequently, for both cases, it is impossible for $i'$, an agent with lower priority than $i^*$, to be matched to $c$ while $i$ is rejected.  

This contradiction invalidates our initial assumption. Therefore, the outcome $\mu$ produced by the MMA algorithm must respect priorities.
\\
\textbf{Maximum Cardinality:}  
In the first step of the MMA algorithm, a maximum size matching $\mu$ is computed. During step 2, if there exists an unmatched agent $i$ with higher priority than a matched agent $i'$ in some category $c$, agent $i$ replaces $i'$ in category $c$, ensuring that the total number of matched agents in the category remains unchanged. If no such unmatched agent $i$ exists, the matching remains unaltered.  
In either case, the number of agents matched to each category remains constant throughout the algorithm. Consequently, the total number of matched agents remains maximal.  
\\
\textbf{Non-wastefulness:}  
Assume, for the sake of contradiction, that the matching produced by the MMA algorithm is not non-wasteful. This would imply the existence of an unmatched agent $i$ and a category $c$ with an unfilled vacancy, where $i$ is eligible for category $c$. In such a case, it would be possible to increase the matching's cardinality by one, contradicting the fact that the algorithm ensures maximum cardinality.  

This completes the proof of Theorem~\ref{theo:MMA:properties}.
\end{proof}


We now present a characterization result for the MMA rule. A matching is said to be \emph{MMA-induced} if it can be generated as an outcome of the MMA rule, starting from some initial maximum matching. Theorem~\ref{theo:characterization:MMA} establishes that a matching satisfies the four fundamental axioms if and only if it is MMA-induced,  derived from the MMA rule with some initial maximum matching.

\begin{restatable}{theorem}{characterizationMMA}
\label{theo:characterization:MMA}
A matching satisfies eligibility compliance, respect for priorities, non-wastefulness, and maximum cardinality if and only if it is MMA-induced. 
\end{restatable}


\begin{proof}
The ``if'' direction has already been established in Theorem~\ref{theo:MMA:properties}. We now prove the ``only if'' direction.

Consider any matching $\mu$ that satisfies the four fundamental properties. To show that $\mu$ can be generated by the MMA algorithm, we slightly modify matching $\mu$ to violate the axiom of respecting priorities by allowing one agent to breach this requirement. Specifically, we choose a category $c$ and an agent $i$ matched to $c$, and then identify an agent $i'$, who i) is unmatched in $\mu$ and ii) has the highest priority among all unmatched agents eligible for category $c$ according to the priority order of category $c$. 

If no such agent exists, the set of maximum matchings must be unique. Otherwise, we obtain a slightly different maximum matching, denoted as $\mu'$ in which $i'$ is matched to category $c$ instead of $i$ while keeping all other agents unchanged. When applying the MMA algorithm to the new matching $\mu'$, we allow agent $i$ to apply for category $c$. Then we will recover the matching $\mu$. 

This completes the proof of Theorem~\ref{theo:characterization:MMA}.
\end{proof}


\begin{theorem}
    \label{theo:MMA:time}
    The MMA algorithm runs in time $O(|E| \sqrt{|V|})$, where $E$ and $V$ denote the number of edges and nodes in the eligibility graph $G$.
\end{theorem}

\begin{proof}
We can compute a maximum matching using the algorithm in \citet{HoKa73a}, which takes time $O(|E| \sqrt{|V|})$. For each unmatched agent $i$, the algorithm checks each eligible category at most once, so updating the matching to respect priorities can be done in time $O(|I| \times |C|)$, which is bounded by $|E|$. Therefore, the total running time is $O(|E| \sqrt{|V|})$.
\end{proof}

Our MMA rule computes a maximum matching only once, with the subsequent adjustment process taking time bounded by $O(|E|)$. In contrast, the REV rule requires computing a maximum matching $|I|$ times, leading to a runtime of $O(|I| \cdot |E| \sqrt{|V|})$. This makes the MMA rule significantly faster than the REV rule by a factor of $|I|$, which is particularly advantageous for large markets, if the primary objective is to satisfy the four fundamental axioms.

\section{Sequentially Processing Reserves}
\label{sec:sequential}


In this section, we introduce a sequential reserve system that extends the basic reserve model in two key ways. First, instead of processing all categories simultaneously, categories can be addressed both sequentially and simultaneously according to a specified precedence order. Second, a subset of categories, termed \emph{preferential treatment categories}, is prioritized to maximize the number of agents matched to them.

The structure of this section is as follows: we first review existing results on a specific model of sequential reserve allocation in healthcare rationing, then introduce a generalized model that builds upon and extends this framework.

\subsection{Previous Results}
A specific form of sequential reserve systems has been previously examined in \citep{PSUY24a, AzBr24a}, where an additional \emph{open (or unreserved)} category, denoted by $c_0$, is introduced alongside the reserved categories. All agents are eligible for this open category $c_0$, with their priority determined by a baseline ranking $\pi$. Categories other than the open category, collectively represented by $C^* = C \setminus \{c_0\}$, are termed \emph{preferential treatment categories}. Agents eligible for each specific category $c \in C$ are known as \emph{beneficiaries}, with the beneficiary group for a particular category $c$ denoted by $I_c \subseteq I$.

We now introduce two axioms that have been proposed for this particular sequential reserve system. The first axiom, presented in Definition~\ref{def:max_beneficial}, defines a matching as \emph{maximal in beneficiary assignment} if the total number of agents assigned to preferential treatment categories is maximized.

\begin{definition}[Maximum Beneficiary Assignment \citep{PSUY24a}]  
\label{def:max_beneficial}  
A matching $\mu \in \Psi$ is called a maximum beneficiary assignment if it maximizes the total number of agents matched to preferential treatment categories. Specifically,  
\[
    \mu \in \arg \max_{\mu' \in \Psi} \sum_{c \in C^*} |\mu'(c)|,  
\]
where $C^*$ represents the set of preferential treatment categories, and $\mu'(c)$ denotes the set of agents matched to category $c$ under matching $\mu'$.
\end{definition}

Definition~\ref{def:max_beneficial} naturally extends to our more general setting, which will be introduced in Section~\ref{sec:sequential:general}. It is important to note that maximum beneficiary assignment remains compatible with maximum cardinality, even when we relax the assumption that all agents must be eligible for the open category $c_0$, a point that will be formally proved in Lemma~\ref{lemma:two_maximum} shortly.

Depending on the order in which the open category $c_0$ and the preferential treatment categories $C^*$ are processed, two types of policies have been explored in various contexts, such as in \citep{Gala84a,HYY13a,DKPS18a}, which dictate the prioritization of categories.

\begin{description}
    \item [Minimum-guarantee policy:] In this approach, the preferential treatment categories are processed first, ensuring that agents eligible for these categories are prioritized before any assignments are made to the open category. This policy aims to guarantee a minimum level of access to preferential treatment categories before considering broader eligibility for the open category.
    \item [Over-and-above policy:] This policy processes the open category first, followed by the preferential treatment categories. It allows agents to be considered for the open category without initially restricting them to any preferential criteria, as long as there are still enough agents qualified for the preferential treatment categories.
\end{description}

\citet{AzBr21a} considered a hybrid policy by dividing the open category $c_0$ into two subcategories, $C_0^1$ and $C_0^2$, and processing them in a specific sequence: first $C_0^1$, then the preferential treatment categories $C^*$, and finally $C_0^2$. This structure allows for flexibility in prioritization. When the capacity of $C_0^2$ is zero, the policy aligns with the minimum-guarantee approach; conversely, when the capacity of $C_0^1$ is zero, it aligns with the over-and-above approach.

For the particular precedence order of $C_0^1$, $C^*$, and $C_0^2$ under the hybrid policy, the authors specifically introduced an axiom called \emph{Order Preservation}, as described in Definition~\ref{def:order_preservation}. Loosely speaking, this axiom stipulates that, given a matching $\mu$, no two agents, who are eligible for both categories they are assigned to in $\mu$, can swap assignments in a way that a higher-priority agent is assigned to a lower-precedence category.

\begin{definition}[Order Preservation \citep{AzBr21a}] 
\label{def:order_preservation}
A matching $\mu$ preserves the precedence order of $C_0^1$, $C^*$, and $C_0^2$ if, for any pair of agents $i, j \in I$ with $i \succ_{\mu(j)} j$, then we have: 
\begin{description}
    \item[(1)] If $\mu(i) \in C^* \cup \{C_0^2\}$ and $j \succ_{\mu(i)} \emptyset$, then $\mu(j) \neq C_0^1$. 
    \item[(2)] If $\mu(j) \in C^* \cup \{C_0^1\}$ and $i \succ_{\mu(j)} \emptyset$, then $\mu(i) \neq C_0^2$. 
\end{description} 
\end{definition}

Formally, Definition~\ref{def:order_preservation} ensures that, for a given matching $\mu$, for any pair of agents $i$ and $j$ where $i$ has higher priority than $j$ for the category $\mu(j)$ assigned to $j$, the following holds: (1) If $i$ is matched to a category in $C^*$ or $C_0^2$ and $j$ is eligible for $i$'s assigned category $\mu(i)$, then $j$ cannot be matched to $C_0^1$. Similarly, (2) if $j$ is matched to a category in $C^*$ or $C_0^1$ and $i$ is eligible for $j$'s assigned category $\mu(j)$, then $i$ cannot be matched to $C_0^2$.

Note that Definition~\ref{def:order_preservation} is specifically defined for the precedence order in the hybrid policy and does not directly extend to our broader sequential reserve model. To address this, we introduce a new axiom in Definition~\ref{def:order_preservation_swap} that generalizes this concept to our setting.

For the hybrid policy, \citet{AzBr24a} proposed the Smart Reverse Rejecting (SREV) rule, which combines the Reverse Rejecting rule (incapable of handling open categories) with the Smart Reserves rule (which cannot accommodate heterogeneous priorities) \citep{SoYe22a}.
Under the hybrid policy, the SREV rule satisfies the four fundamental axioms along with order preservation. However, it does not extend to the more general sequential model introduced in the next subsection.

\begin{proposition}[\citep{AzBr24a}]
    \label{Proposition:hybrid:SREV}
    For the hybrid policy, the Smart Reverse Rejecting (SREV) rule satisfies eligibility compliance, maximum beneficiary assignment, respect for priories, respects for improvements and order preservation.
\end{proposition}

\subsection{Formal Model}
\label{sec:sequential:general}
In this subsection, we introduce a general model for sequential reserve allocation,  by relaxing three key limitations of the hybrid policy \citep{AzBr24a}. First, we allow for any precedence order that accommodates both sequential and simultaneous processing of categories, as opposed to the restrictive order used in the hybrid policy. Second, unlike previous models where open categories must be processed either at the beginning or the end of the sequence, we provide flexibility by permitting open categories to be ``divided'' at any position within the precedence order. Finally, we eliminate the requirement of a baseline ranking for the open category, enabling heterogeneous priorities among agents.

Formally, a sequential reserve system $S = (I, C, q, \succ, C^*, \unrhd)$ extends the basic reserve system $R = (I, C, q, \succ)$ by incorporating two additional components: $C^*$ and $\unrhd$. Specifically, $S$ consists of a set of agents $I$, a set of categories $C$, a capacity vector $q = (q_c)_{c \in C}$, a priority profile $\succ$, a set of preferential treatment categories $C^* \subseteq C$, and a precedence order $\unrhd$ that determines the sequence in which categories are processed.

For any two categories $c, c' \in C$, if $c \unrhd c'$, it means that category $c$ is processed no later than category $c'$, i.e., $c$ has weakly higher precedence than $c'$. If $c \rhd c'$, then category $c$ is processed before category $c'$, indicating that $c$ has strictly higher precedence. If both $c \unrhd c'$ and $c' \unrhd c$ hold, the two categories are tied and are processed simultaneously.

With a slight abuse of notation, we assume that for all categories $c \in C$, $c \rhd \emptyset$. That is, being unmatched is considered as a dummy category with the lowest precedence.

\begin{restatable}{lemma}{LemmaTwoMaximum}
\label{lemma:two_maximum}
Given a sequential reserve system $S$, 
there always exists a matching that satisfies both maximum cardinality and maximum beneficiary assignment.
\end{restatable}

\begin{proof}
First, construct an eligibility graph $G^*$ between agents $I$ and 
preferential treatment categories $C^*$. Compute a maximum cardinality matching in $G^*$, denoted as $\mu$. Next, expand the graph $G^*$ to include all open categories, forming an updated graph $G$. Subsequently, update the matching $\mu$ to maintain maximum size by identifying and augmenting along augmenting paths.

By definition, an augmenting path ensures that every time the matching is updated, all nodes connected to the preferential treatment categories remain matched. Therefore, the number of agents matched to the preferential categories does not decrease during the process. As a result, the compatibility between maximum cardinality and maximum beneficiary assignment is preserved.
\end{proof}

\subsection{Two New Axioms}
Next, we propose two new axioms for the sequential reserve allocation model. The first axiom, \emph{Order Preservation by Swapping}, extends the concept introduced in Definition~\ref{def:order_preservation}. This axiom formalizes the idea that a lower-priority agent $j$ should not occupy an earlier-processed category when a higher-priority agent $i$ could be assigned to that category through a swap, as long as the swap does not violate the conditions of maximum cardinality and maximum beneficiary assignment.

\begin{definition}[Order Preservation by Swapping]
\label{def:order_preservation_swap}
Given a sequential reserve system $S$, a matching $\mu$ satisfies \emph{Order Preservation by Swapping} with respect to the precedence order $\unrhd$ if no pair of agents $i, j \in I$ exists such that:
    \begin{itemize}
        \item $\mu(j) \rhd \mu(i)$, i.e., category $\mu(j)$ is processed before category $\mu(i)$;
        \item $i \succ_{\mu(j)} j$, i.e., agent $i$ has a higher priority than agent $j$ for category $\mu(j)$;
        \item $j \succ_{\mu(i)} \emptyset$ and $i \succ_{\mu(j)} \emptyset$, i.e., both agents are eligible for each other's assigned categories.
    \end{itemize}
\end{definition}

One can verify that the axiom in Definition~\ref{def:order_preservation_swap} captures the same principle as Definition~\ref{def:order_preservation} under the specific hybrid policy, with the important distinction that Definition~\ref{def:order_preservation_swap} can be applied to any precedence order $\unrhd$.

\begin{proposition}\label{prop:order_preservation:equivalence}
    For the specific order under the hybrid policy, Definition~\ref{def:order_preservation_swap} becomes equivalent to Definition~\ref{def:order_preservation}.
\end{proposition}

Next, we introduce a new axiom called \emph{Respect for Precedence over Categories}, which is stronger than the notion of \emph{Order Preservation by Swapping}. The intuition is to ensure that for a given precedence order $\rhd$ over categories, a category $c$ is always matched with a set of agents with higher priority, provided i) this does not alter the assignment of any categories processed prior to $c$, and ii) maximum cardinality and maximum beneficiary assignment can be achieved.

The key distinction from \emph{Order Preservation by Swapping} is that this axiom permits reallocations beyond simple swaps between two agents. We illustrate this difference in Example~\ref{example:precedence}.

\begin{example}[Comparison of  Definition~\ref{def:order_preservation_swap} and~\ref{def:respect_precedence}]
    \label{example:precedence}
    Suppose there are three agents $I = \{i_1, i_2, i_3\}$ and two categories $C = \{c_1, c_2\}$, each with a capacity of 1. The precedence order $\rhd$ specifies $c_1$ before $c_2$, and the priority rankings for the categories are as follows:  $i_3 \succ_{c_1} i_1$ and $i_3 \succ_{c_2} i_2$.
    
    Consider the matching $\mu_1$ where $\mu_1(i_1) = c_1$ and $\mu_1(i_3) = c_2$. This matching satisfies \emph{Order Preservation by Swapping}, because agent $i_2$ cannot swap his assignment with agent $i_3$ due to his lower priority for category $c_2$ by Definition~\ref{def:order_preservation_swap}.
    
    An alternative matching $\mu_2$ where $\mu_2(i_3) = c_1$ and $\mu_2(i_2) = c_2$ is more preferable because category $c_1$ is processed first and it is matched with agent $i_3$ who holds the highest priority. However, we cannot obtain this new matching from $\mu_1$ through a simple swap of assignments between $i_2$ and $i_3$. Instead, it is necessary to adjust the assignments of the three agents.
\end{example}

Next, we introduce the concept of Respect for Precedence over Categories. 

\begin{definition}[Respect for Precedence over Categories]
    \label{def:respect_precedence}
    A matching $\mu$ respects precedence over categories with respect to $\rhd$ if there does not exist a pair of agents $i, j \in I$ such that:
    \begin{description}
        \item [1) Higher Precedence and Lower Priority:] $\mu(j) \rhd \mu(i)$ and $i \succ_{\mu(j)} j \succ_{\mu(j)} \emptyset$;
        \item [2) Existence of an Alternative Matching:] There exists an alternative matching $\mu'$ such that:
        \begin{description}
            \item [i)] For all $c' \rhd \mu(j)$, and for all $k \in \mu(c')$, we have $\mu(k) = \mu'(k)$;
            \item [ii)] For all $\ell \in \mu(j)$ with $\ell \succ_{\mu(j)} i$, we have $\mu(\ell) = \mu'(\ell)$;
            \item [iii)] $\mu'(i) = \mu(j)$;
            \item [iv)] The matching $\mu'$ still satisfies the conditions of maximum cardinality and maximum beneficiary assignment.
        \end{description}
    \end{description}
\end{definition}

Definition~\ref{def:respect_precedence} states that there does not exist a pair of agents $i$ and $j$ satisfying the following two conditions. 

First, the 
category $\mu(j)$ is processed prior to category $\mu(i)$, and agent $j$ has a lower priority than agent $i$ for category $\mu(j)$. In other words, 
from the perspective of category $\mu(j)$, it has a higher precedence than category $\mu(i)$ according to $\rhd$, but it is matched with some agent $j$ who has a lower priority than some agent $i$ who is matched to a later processed category $\mu(i)$.

Second, there exists an alternative matching $\mu'$ such that (i) for any category $c' \rhd \mu(j)$ processed prior to $\mu(j)$, every agent matched to $c'$ in $\mu$ remains matched to the same category in $\mu'$; (ii) for any agent $\ell$ matched to category $\mu(j)$ with a higher priority than agent $i$ in $\mu$, the assignment for agent $\ell$ remains unchanged in $\mu'$; (iii) agent $i$ is matched to category $\mu(j)$ in the matching $\mu'$; and (iv) the matching $\mu'$ still satisfies the requirement of maximum cardinality and maximum beneficiary assignment. 

\begin{restatable}{proposition}{propConnectionPrecedence}\label{prop:connection:precedence}
    Respect for precedence over categories implies order preservation by swapping.
\end{restatable}


\begin{proof}
    Suppose, for contradiction, that there exists a pair of agents $i, j \in I$ that violates order preservation by swapping. That is, the following conditions hold:
    \begin{itemize}
        \item $\mu(j) \rhd \mu(i)$, meaning category $\mu(j)$ is processed before category $\mu(i)$;
        \item $i \succ_{\mu(j)} j$, meaning agent $i$ has a higher priority than agent $j$ for category $\mu(j)$;
        \item $j \succ_{\mu(i)} \emptyset$ and $i \succ_{\mu(j)} \emptyset$, meaning both agents are eligible for each other's assigned categories.
    \end{itemize}

    Since both agents are eligible for each other's assigned categories and $i$ has a higher priority than $j$ for category $\mu(j)$, there exists an alternative matching $\mu'$ where $i$ and $j$ swap their assignments, i.e., $\mu'(i) = \mu(j)$ and $\mu'(j) = \mu(i)$. Since the assignments of all other agents remain unchanged, this contradicts the requirement of respect for precedence over categories. 

    Therefore, any matching that violates order preservation by swapping must also violate respect for precedence over categories. This establishes that respect for precedence over categories implies order preservation by swapping.

    This completes the proof of Proposition~\ref{prop:connection:precedence}.
\end{proof}

\section{Sequential Category Updating Rule}
\label{sec:SCU}

In this section, we introduce the Sequential Category Updating (SCU) Rule, as detailed in Algorithm~\ref{alg:SCU:rule}. This rule accommodates both the basic reserve system and the sequential reserve system. It not only satisfies all the axioms and properties introduced in Section~\ref{sec:Preliminaries} (Theorem~\ref{theorem:SCU:properties}), but is also the unique rule that satisfies the key axioms under a strict precedence order over categories (Theorem~\ref{theorem:SCU:unique}).


The Sequential Category Updating (SCU) rule starts by computing an initial matching $\mu$ that satisfies maximum cardinality and maximum beneficiary assignment. Throughout the process, the algorithm maintains two sets: $X$, which contains agents whose assignments have been finalized, and $Y$, which includes categories that have already been processed.

A strict precedence order $\rhd$ is derived from the given precedence order $\unrhd$ by breaking ties based on smaller category indices. Using this derived order, the algorithm iterates over the categories, starting with the highest-ranked category.
For each category $c$, the algorithm considers eligible agents who are not yet assigned, sorted in descending priority order according to $\succ_c$. For each eligible agent $i$, it checks whether a modified matching $\mu'$ can be found that:
\begin{enumerate}
    \item Preserves the current assignments of all agents in $X$.
    \item Matches $i$ to category $c$.
    \item Satisfies maximum cardinality and maximum beneficiary assignment.
\end{enumerate}
If such a matching $\mu'$ exists, the algorithm updates $\mu$ to $\mu'$, adds $i$ to $X$, and proceeds. This ensures that all updates preserve the persistency of existing assignments and respect the priority ordering. The process repeats until all categories have been processed, yielding a final matching $\mu$.

\begin{algorithm}[tb]
\caption{Sequential Category Updating (SCU) Rule}
\label{alg:SCU:rule}

\KwIn{A sequential reserve system $S = (I, C, C^*, q, \unrhd)$}
\KwOut{A matching $\mu$}

Compute a matching $\mu$ satisfying both maximum cardinality and maximum beneficiary assignment\;
Initialize $X \gets \emptyset$ to store the set of agents whose assignments are determined\;
Initialize $Y \gets \emptyset$ to store the set of categories that have been processed\;
Derive a strict order $\rhd$ from $\unrhd$ by breaking ties based on smaller category indices\;

\ForEach{$c \in C \setminus Y$ in descending order of $\rhd$}{
    \While{$|X| \cap \mu(c) < q_c$ and there exists an eligible agent $i \in I_c \setminus X$ in descending order of $\succ_c$}{
        \If{there exists a matching $\mu'$ such that:
        \begin{enumerate}
            \item $\mu'(i') = \mu(i')$ for all $i' \in X$,
            \item $\mu'(i) = c$, and
            \item $\mu'$ satisfies maximum cardinality and maximum beneficiary assignment
        \end{enumerate}}{
            Update $\mu \gets \mu'$ and $X \gets X \cup \{i\}$\;
        }
    }
    Update $Y \gets Y \cup \{c\}$\;
}
\Return{The final matching $\mu$}
\end{algorithm}

\subsection{Properties of SCU}


To facilitate the analysis of the proofs, we first introduce a new notion.

\begin{definition}
    \label{def:addition_position} 
    Let $X$ denote the set of agents whose assignments are determined during the execution of the SCU rule. For any agent $i$, let $Pos(X, i)$ represent the position at which agent $i$ is added to $X$. If an agent $j$ is never added to $X$, we define $Pos(X, j) = \infty$.
\end{definition}

The following lemma states that, given the matching $\mu$ yielded by the SCU rule, for two agents $i$ and $j$, where $i$ has higher priority than $j$ at the category $\mu(j)$, if the assignment of agent $i$ is determined after that of agent $j$ under the SCU rule, then we have that: (1) agent $i$ must be matched to a category $c' \in C$ different from $\mu(j)$, and (2) all agents eligible for category $c'$ are matched in the final matching $\mu$.
 Formally, 

\begin{restatable}{lemma}{LemmaLaterMatched}
\label{lemma:later_matched}
    Let $\mu$ denote the matching returned by SCU. Consider any two agents $i$ and $j$. If the following conditions hold:
    \begin{itemize}
        \item $i \succ_{\mu(j)} j \succ_{\mu(j)} \emptyset$, i.e., agent $i$ has higher priority than agent $j$ at the category $\mu(j)$,
        \item $Pos(X, i) > Pos(X, j)$, i.e., the assignment of agent $i$ is determined after that of agent $j$,
    \end{itemize}
    then we have:
    \begin{description}
        \item[(i)] $\mu(i) \in C \setminus \{\mu(j)\}$, i.e., agent $i$ is matched to a different category.
        \item[(ii)] All agents eligible for category $\mu(i)$ are matched in the final matching $\mu$.
    \end{description}
\end{restatable}


\begin{proof}
 We first prove part (i), that is, agent $i$ must be matched to some category $c' \in C$ different from $\mu(j)$. For the sake of contradiction, assume that agent $i$ is unmatched in the final matching $\mu$. Consider a different matching $\mu'$ in which agent $i$ is matched to category $c$, while agent $j$ is unmatched. In other words, let agent $i$ replace agent $j$ for category $c$. Note that the number of agents matched to each category does not change, and all agents except for $i$ and $j$ remain unchanged. Since this results in a valid matching that satisfies the two maximum properties (maximum cardinality and maximum beneficiary assignment), and all agents whose assignments were fixed before checking $i$ remain unchanged, SCU would have matched $i$ to category $\mu(j)$ instead, leading to a contradiction that $i$ is unmatched in $\mu$. Therefore, agent $i$ must be matched in the final matching.

Next, we prove part (ii), that is, agent $i$ must be matched to a category $c'$, for which all agents who are eligible for $c'$ are also matched in $\mu$. For the sake of contradiction, assume that there exists an agent $\ell \in I_{c'}$ who is unmatched in the output matching $\mu$. Then modify the matching $\mu$ to derive a new matching $\mu'$ where agent $i$ is matched to $c$, agent $j$ becomes unmatched, agent $\ell$ is matched to $c'$, and all other agents' assignments remain unchanged. Since this matching satisfies the two maximum properties and all agents whose assignments were determined before checking agent $i$ remain unchanged, SCU would have matched $i$ to category $\mu(j)$ instead, leading to a contradiction that $i$ is matched to $\mu(i)$. Therefore, no such unmatched agent $\ell$ can exist, and all agents eligible for category $c'$ must be matched in the final outcome.

This completes the proof of Lemma~\ref{lemma:later_matched}.
\end{proof}

The implication of Lemma~\ref{lemma:later_matched} is that if an agent $i$ is not matched to an eligible category $c$, but another agent $j$ with a lower priority is matched to $c$, then agent $i$ must be matched to some other category $c'$, where agent $i$ plays a critical role in achieving maximum cardinality or maximum beneficiary assignment.

The following lemma states that all agents who are matched in the final matching yielded by the SCU rule are added one by one to the set $X$, which stores the set of agents whose assignments are finalized during the execution of the SCU rule. In other words, the set $X$ consists exactly of all the agents matched by SCU.

\begin{restatable}{lemma}{LemmaEqualSet}
\label{lemma:equal_set}
Let $\mu$ denote the matching yielded by the SCU rule, and let $\mu(C)$ denote the set of agents matched to some category $c \in C$. Then, we have $X = \mu(C)$, meaning that the set $X$ of agents whose assignments are determined during the execution of SCU is exactly the set of agents matched in the final matching.
\end{restatable}


\begin{proof}
We first prove that $X \subseteq \mu(C)$. Recall that the set of chosen agents in $X$ are added one by one during the algorithm's execution. Consider an agent $i \in X$. Agent $i$ is added to $X$ if there exists a matching that satisfies two properties (maximum cardinality and maximum beneficiary assignment) and does not affect any agents matched previously. Therefore, once $i$ is added, it stays matched to the same category until the algorithm terminates. This ensures that for each $i\in X$, we have
$i \in \mu(C)$, and thus we can infer that $X \subseteq \mu(C)$.

Next, we prove that $\mu(C) \subseteq X$. Assume, for the sake of contradiction, that there exists an agent $i \in \mu(C) \setminus X$. Let $C_i$ denote the set of categories for which agent $i$ is eligible. When processing each category $c \in C_i$, the algorithm checks all eligible agents $I_c$ and does not add $i$ to $X$. This indicates that there is no matching where i) agents processed before $i$ remain matched to the same category, and ii) the matching satisfies the two properties (maximum cardinality and maximum beneficiary assignment). However, this contradicts the fact that the final outcome $\mu$ satisfies these properties. Thus the assumption is wrong and therefore, then we have $\mu(C) \subseteq X$.

Since $X \subseteq \mu(C)$ and $\mu(C) \subseteq X$ hold, we know $X=\mu(C)$, completing the proof of Lemma~\ref{lemma:equal_set}.
\end{proof}

We apply Lemma~\ref{lemma:later_matched} and Lemma~\ref{lemma:equal_set} to prove Theorem~\ref{theorem:SCU:properties} and Theorem~\ref{theorem:SCU:unique}.





Theorem~\ref{theorem:SCU:properties} states that the SCU rule satisfies not only the four fundamental axioms but also the new axioms for sequential reserve allocation. In addition, it satisfies two incentive properties and guarantees a consistent set of matched agents.

\begin{restatable}{theorem}{theoremSCUproperties}
\label{theorem:SCU:properties}  
Given a sequential system $S$, the Sequential Category Updating (SCU) rule satisfies the following properties: 
\begin{description}
    \item[i) Four Fundamental Axioms: ] eligibility compliance, maximum cardinality, non-wastefulness, and respect for priorities;
    \item[ii) Three Sequential Axioms: ] maximum beneficiary assignment, order preservation by swap, and respect for precedence over categories;
    \item[iii) Two Incentive Properties: ] no incentive to hide and respect for improvements;
    \item[iv) Three Consistency Properties: ] independence of baseline, consistent matched agents and consistent matching.
\end{description}  
\end{restatable}

\begin{proof}  
The proof proceeds by verifying each property:  
\\
\textbf{Eligibility Compliance:}  
The SCU rule sequentially examines eligible agents for each category according to the precedence ordering $\rhd$. An agent $i$ can only be matched to a category if $i$ qualifies for it. Therefore, the SCU rule guarantees that no category is ever matched with an ineligible agent, ensuring eligibility compliance.
\\
\textbf{Maximum Cardinality and Maximum Beneficiary Assignment:}  
The SCU algorithm begins with a matching that satisfies both maximum cardinality and maximum beneficiary assignment. Each time an agent $i$ is added to the matching $X$, we ensure that there exists a matching in which $i$ is assigned while maintaining both of these maximum properties. As the algorithm progresses, these two properties are preserved. When the algorithm terminates, the final matching satisfies both maximum cardinality and maximum beneficiary assignment.
\\
\textbf{Non-Wastefulness:}  
Maximum cardinality inherently ensures non-wastefulness, as the matching guarantees that no eligible agent is left unmatched when there are vacancies in any category for which they qualify.
\\
\textbf{Respect for Priorities:}  
Let $\mu$ denote the final outcome matching. For the sake of contradiction, suppose there exists a pair of agents $i$ and $j$ such that $\mu(j) = c$, $\mu(i) = \emptyset$, and $i \succ_{\mu(j)} j$. By Lemma~\ref{lemma:later_matched}, we know that agent $i$ cannot be unmatched, leading to a contradiction. Thus, the assumption is incorrect, and the matching $\mu$ satisfies respect for priorities.
\\
\textbf{Respect of Precedence Order over Categories:} 
For contradiction, assume that in the matching $\mu$ produced by the SCU algorithm, there exists a pair of agents $i, j \in I$ such that:
\begin{itemize}
    \item $\mu(j) \rhd \mu(i)$ and $i \succ_{\mu(j)} j \succ_{\mu(j)} \emptyset$;
    \item There exists an alternative matching $\mu'$ such that:
    \begin{description}
        \item [i)] For all $c' \rhd \mu(j)$, and for all $k \in \mu(c')$, we have $\mu(k) = \mu'(k)$;
        \item [ii)] For all $\ell \in \mu(j)$ with $\ell \succ_{\mu(j)} i$, we have $\mu(\ell) = \mu'(\ell)$;
        \item [iii)] $\mu'(i) = \mu(j)$;
        \item [iv)] The matching $\mu'$ still satisfies the conditions of maximum cardinality and maximum beneficiary assignment.
    \end{description}
\end{itemize}
Consider a different matching $\mu'$ in which agent $i$ is matched to category $c$ while agent $j$ is unmatched, i.e., let $i$ replace $j$ for category $c$. Note that the number of agents matched to each category does not change, and all agents except for $i$ and $j$ remain unchanged. Thus, there exists another matching $\mu'$ that satisfies both maximum cardinality and maximum beneficiary assignment. Since the assignments of agents fixed before checking $i$ remain unchanged, this leads to a contradiction. 
\\
\textbf{Order Preservation by Swap:} By Proposition~\ref{prop:connection:precedence}, respect for the precedence order over categories implies order preservation by swap. Therefore, the SCU rule satisfies this property as well.
\\
\textbf{Respect Improvements:} Consider any agent $i$ who is matched to some category $c$ under the SCU rule. Suppose we improve $i$'s priority for some category $c'$, while the priorities of all other agents remain unchanged.
If $c \rhd c'$, then the outcome does not change.
Suppose $c' \rhd c$. If agent $i$'s updated priority is higher than that of agent $i'$, who has the lowest priority matched to category $c'$, then by Lemma~\ref{lemma:later_matched}, we know that agent $i$ cannot be unmatched. Otherwise, agent $i$ remains matched to category $c$.
\\
\textbf{No Incentive to Hide:} By Proposition~\ref{proposition:connection:manipulation}, we know that Respect Improvements implies No Incentive to Hide. Thus, the SCU rule satisfies the latter property.
\\
\textbf{Consistent Matching:} Suppose the SCU rule does not return the same matching for the same input instance. Let $i$ denote the first agent who receives a different assignment under two runs of the SCU rule. Suppose $i$ is matched to $c$ in one matching $\mu$ and is matched to $c'$ in the other matching $\mu'$. Furthermore, suppose $c \rhd c'$, i.e., $c$ is processed prior to $c'$. 
By the assumption that $i$ is the first agent to receive a different assignment, we know that no agent matched to $c'' \rhd c$ has its assignment changed. Now, consider the second run of SCU in which $i$ is matched to $c'$. Since $i$ was assigned to $c$ in the first run, we know there exists a matching $\mu$ in which $i$ can be matched to $c$ along with all agents who are added to $X$ prior to $i$. This leads to a contradiction, as it would imply that the SCU does not match $i$ to $c$ in the second run, contradicting the assumption that $i$ is the first agent with a different assignment. Hence, the SCU rule must return the same matching for the same input instance.
\\
\textbf{Consistent Matched Agents:} Since consistent matching implies consistent matched agents, it follows that SCU satisfies the latter property.
\\
\textbf{Independence of Baseline:} The SCU rule does not rely on a baseline priority order over agents.

This completes the proof of Theorem~\ref{theorem:SCU:properties}.
\end{proof}  


Theorem~\ref{theorem:SCU:unique} states that given a strict precedence order over categories, the SCU rule is the unique one that satisfies eligibility compliance, maximum cardinality, non-wastefulness, respect for priorities, maximum beneficiary assignment and respect for precedence.

\begin{restatable}{theorem}{theoremSCUunique}
\label{theorem:SCU:unique}  
Given a sequential system $S$ with a strict precedence order $\rhd$, an algorithm satisfies eligibility compliance, maximum cardinality,  maximum beneficiary assignment, and respect of precedence if and only if it is the Sequential Category Updating (SCU) rule. Furthermore, the matching yielded by the SCU rule is unique.
\end{restatable}  

\begin{proof}
Consider any instance of a sequential reserve system, and let $\mu$ denote the outcome returned by the SCU rule.  For the sake of contradiction, suppose there exists another rule $\Gamma$ that satisfies eligibility compliance, non-wastefulness, maximum cardinality, maximum beneficiary assignment, and respect of precedence, but returns an outcome $\mu'$. 

Next we prove that $\mu = \mu'$, which implies that not only the set of matched agents $\mu'(C)$ is the same as $\mu(C)$, but also each agent is matched to the same category in two matchings.

Suppose we are given a strict precedence order over categories, denoted as $c_1, c_2, \dots, c_k$.  
Consider the set of agents matched to $c_1$ in $\mu$, denoted as $\mu(c_1)$. Let $i^* \in \mu(c_1)$ be the agent with the lowest priority among those in $\mu(c_1)$. Then, we can divide all agents $I_{c_1}$ who are eligible for $c_1$ into three disjoint groups as follows:  
\begin{itemize}
    \item $O = \mu(c_1)$: the set of agents who are matched to category $c_1$ in the matching $\mu$.
    \item $P = \{i \in I_{c_1} \mid i \succ_{c_1} i^* \text{ and } i \notin \mu(c_1)\}$: the set of agents who have higher priority than $i^*$ but are not matched to category $c_1$.
    \item $Q = I_{c_1} \setminus (O \cup P)$: the set of agents with lower priority than $i^*$.
\end{itemize}

Check each agent $i \in I_{c_1}$ in descending order of $\succ_{c_1}$ as follows:
\begin{itemize}
    \item If agent $i \in O$, then $i$ must be matched to $c_1$ in $\mu'$. Otherwise, matching $\mu'$ cannot satisfy respect of precedence, as agent $i$ can be matched to $c_1$ along with all higher-priority agents in $O$ in a matching that satisfies maximum cardinality and maximum beneficiary assignment.
    \item If agent $i \in P$, then $i$ cannot be matched to $c_1$ in $\mu'$. Otherwise, matching $\mu'$ cannot satisfy maximum cardinality or maximum beneficiary assignment. This is because any agent with a higher priority than $i^*$ but not included in $\mu(c_1)$ must be assigned to a different category by Lemma~\ref{lemma:later_matched}. 
    \item If agent $i \in Q$, then $i$ cannot be matched to $c_1$. Otherwise, there must exist some agent $i' \in O$ who is not matched to $c_1$, which means that $\mu'$ cannot satisfy respect of precedence, as agent $i'$ can replace agent $i$.
\end{itemize}
Thus, we have for each $i\in I_{c_1}$
$\mu(i) = \mu'(i)$, and all agents in $O$
are matched to $c_1$.

Now, consider category $c_k$ for any $k \geq 2$. 
By induction, we apply the same argument after removing $\mu(I_{c_{k-1}})$ and their assignments from consideration. 
For each category $c$, we have $i\in I_{c}$
$\mu(i) = \mu'(i)$, which completes the proof of Theorem~\ref{theorem:SCU:unique}.
\end{proof}

\section{Implementation of The Sequential Category Updating Rule}
\label{sec:SCU:implementation}

In this section, we present two implementations of the Sequential Category Updating (SCU) rule based on two graph structures: flow networks and bipartite graphs. While previous work \citep{AzBr21a, AzBr24a, PSUY21a, PSUY24a} primarily utilized bipartite graphs, we show that implementing the rule using flow networks offers a more intuitive approach and could significantly improve computational efficiency when grouping agents based on their eligible categories. For completeness, we also provide a fast implementation of the SCU rule using bipartite graphs.

\subsection{Implementation Based on Flow Network}

A flow network $F = (V, E)$ is a directed graph where each edge $(v, w) \in E$ carries a flow subject to a lower bound $\ell(v, w)$ and an upper bound $u(v, w)$. The network also includes a source vertex $s$ and a sink vertex $t$.

\begin{definition}[Flow]
\label{def:flow}
A \emph{flow} in a network $F$ is a function $f: V \times V \to \mathbb{R}$ that satisfies the following conditions:
\begin{enumerate}
    \item \textbf{Capacity Constraint}: For every edge $(v, w) \in E$,
    \[
        \ell(v, w) \leq f(v, w) \leq u(v, w),
    \]
    where $\ell(v, w)$ and $u(v, w)$ denote the lower and upper bounds of the edge $(v, w)$, respectively.
    
    \item \textbf{Flow Conservation}: For every vertex $w \in V \setminus \{s, t\}$,
    \[
        \sum_{v \in V} f(v, w) = \sum_{v \in V} f(w, v).
    \]
    That is, the total incoming flow to $w$ must equal the total outgoing flow from $w$.
\end{enumerate}
\end{definition}

Given a sequential reserve system $S = (I, C, q, \succ, C^*, \unrhd)$, we construct a corresponding flow network $F = (V, E)$, which consists of three main layers:  

\begin{description}
    \item [First Layer:] This layer contains $|I|$ nodes, each representing an agent $i \in I$. A directed edge connects the source vertex $s$ to each agent node $i$, with a lower bound of $0$ and an upper bound of $1$. A flow value of $f(s, i) = 1$ indicates that agent $i$ is matched.
      
    \item [Second Layer:] This layer consists of $|C|$ nodes, each corresponding to a category $c \in C$. For every eligible agent $i \in I_c$ within category $c$, a directed edge is created from agent node $i$ in the first layer to category node $c$ in the second layer. Each edge $(i, c)$ has an initial lower bound of $0$ and an upper bound of $1$. The lower bound is updated during the algorithm and a flow value of $f(i, c) = 1$ indicates that agent $i$ is matched to category $c$.  
      
    \item [Third Layer:] This layer includes two nodes, $C_0$ and $C^*$, representing open and beneficial categories, respectively.  
    \begin{itemize}
        \item For each beneficial category $c^* \in C^*$, an edge is added from $c^*$ in the second layer to $C^*$ in the third layer, with a lower bound of $0$ and an upper bound of $q_{c^*}$.  
        \item For each open category $c' \in C \setminus C^*$, an edge is added from $c'$ in the second layer to $C_0$ in the third layer, with a lower bound of $0$ and an upper bound of $q_{c'}$.  
        \item Finally, directed edges connect both $C_0$ and $C^*$ to the sink vertex $t$, with an upper bound of $q$. The lower bound is determined during the algorithm.  
    \end{itemize}
\end{description}

\begin{algorithm}[tb]
    \caption{Implementation of the Sequential Category Updating Rule Using a Flow Network}
    \label{algorithm:implementation:flow}
    \SetAlgoLined
    \KwIn{A sequential reserve system $(I, C, q, \succ, C^*, \unrhd)$}
    \KwOut{A feasible flow $f$ representing the allocation of agents to categories.}
    
    Construct and initialize a flow network $F$ with all lower bounds set to zero\;
    
    \textbf{Step 1: Compute Maximum Beneficiary Assignment} \\
    Set the upper bound of the edge $(C_0, t)$ to zero\;
    Compute a maximum flow and record its total value as $b$\;
    
    \textbf{Step 2: Compute Maximum Cardinality} \\
    Set the upper bound of the edge $(C_0, t)$ to $q$\;
    Set the lower bound of the edge $(C^*, t)$ to $b$\;
    Compute a maximum flow $f$ and record its total value as $m$\;
    
    \textbf{Step 3: Sequential Updating Procedure} \\
    Set the lower bound of the edge $(C_0, t)$ to $m - b$\;
    Set the lower bound of the edge $(C^*, t)$ to $b$\;
    
    Initialize $X \leftarrow \emptyset$\;
    \ForEach{category $c$ in descending order of $\rhd$ (breaking ties by smaller indices)}{
        \ForEach{agent $i \in I_c \setminus X$ in descending order of $\succ_c$}{
            \If{a feasible flow exists when setting the lower bound of edge $(i, c)$ to $1$}{
                Set the lower bound of the edge $(i, c)$ to $1$\;
                Add $i$ to $X$\;
            }
        }
    }
    \Return a flow $f$, where $f(i, c) = 1$ indicates agent $i$ is assigned to category $c$\;
\end{algorithm}

We now describe the implementation of the SCU rule using a flow network, as formalized in Algorithm~\ref{algorithm:implementation:flow}. First, we compute a maximum flow by setting the upper bound of the edge $(C_0, t)$ to zero. This prevents any flow from passing through that edge. The total flow value from this computation, denoted as $b$, represents the maximum number of agents matched to beneficial categories.
Next, we compute a second maximum flow by setting the upper bound of the edge $(C_0, t)$ to $q$ and the lower bound of the edge $(C^*, t)$ to $b$. This ensures that the total number of agents matched to all categories is maximized, while maintaining the maximum number of agents matched to beneficial categories. The total flow value from this computation, denoted as $m$, represents the maximum number of agents matched to all categories.

After obtaining $b$ and $m$, we update the flow network by setting the lower bound of the edge $(C_0, t)$ to $m - b$ and the lower bound of the edge $(C^*, t)$ to $b$. The final assignment is determined through an iterative procedure. We initialize $X \leftarrow \emptyset$ (the set of agents whose assignments are finalized) and process each category $c$ in the precedence order defined by $\rhd$. For each agent $i \in I_c \setminus X$, ordered according to $\succ_c$, we check for a feasible flow when setting the lower bound of the edge $(i, c)$ to 1. If a feasible flow exists, we assign $\ell(i, c) = 1$ (indicating that $i$ is matched to $c$) and add $i$ to $X$; otherwise, we move on to the next agent.

\begin{restatable}{theorem}{theoremImplementationFlow}
    \label{theorem:implementation:flow}
    Given a sequential reserve system $S$ and its corresponding flow network $F$, Algorithm~\ref{algorithm:implementation:flow} computes a flow $f$ that corresponds to the assignment produced by the SCU rule in time $O(|I| \cdot |C| \cdot |V|^3)$ where $|V|$ denotes the number of nodes in $F$.
\end{restatable}

\begin{proof}
In steps 1 and 2 of Algorithm~\ref{algorithm:implementation:flow}, we first compute a flow that maximizes the number of agents assigned to beneficial categories, with a total flow value denoted as $b$. Next, we compute a flow that maximizes the overall number of matched agents, with a total flow value denoted as $m$. These steps correspond to constructing a matching $\mu$ in the SCU rule that satisfies both the maximum cardinality and maximum beneficiary assignment conditions.

The core procedure of the SCU rule is to determine whether an agent $i$ can be assigned to category $c$ while preserving the assignments of agents in $X$. Specifically, the SCU rule verifies whether there exists an alternative matching $\mu'$ in which agent $i$ is matched to category $c$, all agents in $X$ retain their assignments, and $\mu'$ continues to satisfy both the maximum cardinality and maximum beneficiary assignment conditions.

In the flow network, this verification is straightforward. We enforce a lower bound of $b$ on the edge $(C^*, t)$ and $m - b$ on the edge $(C_0, t)$ to ensure that the number of agents matched to beneficiary and open categories remains maximal. For each agent $j \in X$, if they are assigned to a category $c'$, we impose a lower bound of $1$ on the edge $(j, c')$ to guarantee that their assignment remains unchanged. Next, we check whether agent $i$ can be matched to category $c$ by setting a lower bound of $1$ on the edge $(i, c)$. If a feasible flow exists under these constraints, it confirms that agent $i$ can be assigned to category $c$ without altering the assignments of agents in $X$, while still satisfying both the maximum cardinality and maximum beneficiary assignment conditions.

This completes the proof of Theorem~\ref{algorithm:implementation:flow}.
\end{proof}

\begin{algorithm}[tb]
    \caption{Implementation of the SCU Rule Using a Compact Flow Network}
    \label{algorithm:implementation:flow:compact}
    \SetAlgoLined
    \KwIn{A sequential reserve system $(I, C, q, \succ, C^*, \unrhd)$.}
    \KwOut{An allocation $\mu$ of agents to categories.}
   
    Construct and initialize a compact flow network $F^*$ with all lower bounds set to zero\;
    
    \textbf{Step 1: Compute Maximum Beneficiary Assignment} \\
    Set the upper bound of edge $(C_0, t)$ to zero\;
    Compute a maximum flow and record its total value as $b$\;
    
    \textbf{Step 2: Compute Maximum Cardinality} \\
    Set the upper bound of edge $(C_0, t)$ to $q$\;
    Set the lower bound of edge $(C^*, t)$ to $b$\;
    Compute a maximum flow $f$ and record its total value as $m$\;
    
    \textbf{Step 3: Sequential Updating Procedure} \\
    Set the lower bound of edge $(C_0, t)$ to $m - b$\;
    Set the lower bound of edge $(C^*, t)$ to $b$\;
    
    Initialize $X \leftarrow \emptyset$ (set of assigned agents) and $\mu \leftarrow \emptyset$ (final allocation)\;
    
    \ForEach{category $c$ in descending order of $\rhd$ (breaking ties by smaller indices)}{
        \ForEach{agent $i \in I_c \setminus X$ in descending order of $\succ_c$}{
            \If{a feasible flow exists when setting the lower bound of edge $(k, c)$ to $1$, where $i$ belongs to group $k$}{
                Assign $i$ to category $c$, i.e., set $\mu(i) = c$\;
                Add $i$ to $X$\;
                Decrease the upper bound of edge $(k, c)$ by $1$\;
            }
        }
    }
    
    \Return allocation $\mu$\;
\end{algorithm}

We next present a more compact flow network, where all agents are partitioned into distinct groups $K$ based on the set of categories they are eligible for. The key difference is that the first-layer nodes are replaced by $K$, each representing a distinct group. 
Formally, given a sequential reserve system $S = (I, C, q, \succ, C^*, \unrhd)$, we construct a compact flow network $F^* = (V, E)$, which consists of three main layers:

\begin{description}
    \item [First Layer:] This layer contains $|K|$ nodes, each representing a distinct group instead of individual agents. A directed edge connects the source vertex $s$ to each group node $k$, with a lower bound of $0$ and an upper bound of $|I_k|$, where $I_k$ is the set of agents belonging to group $k$.
      
    \item [Second Layer:] This layer consists of $|C|$ nodes, each corresponding to a category $c \in C$. For every node $k \in K$ in the first layer, a directed edge is created from group node $k$ to category node $c$ in the second layer if $c$ is associated with group $k$.
      
    \item [Third Layer:] This layer remains unchanged; all nodes and edges are identical to those in the original flow network.
\end{description}

We can slightly modify Algorithm~\ref{algorithm:implementation:flow} without altering the first two steps. Specifically, in the third step, whenever an agent $i$ is selected to be added to $X$, we assign $i$ to the corresponding category $c$ in the final assignment $\mu$. However, instead of continuing to track $i$ individually, we remove it from consideration and decrement the upper bound of its group $k$ by $1$. The formal description of this modified algorithm is provided in Algorithm~\ref{algorithm:implementation:flow:compact}.

\begin{restatable}{theorem}{theoremImplementationFlowCompact}
    \label{theorem:implementation:flow:Compact}
    Given a sequential reserve system $S$ and its corresponding compact flow network $F^*$, Algorithm~\ref{algorithm:implementation:flow:compact} computes a flow $f$ that corresponds to the assignment produced by the SCU rule in time $O(|I| \cdot |C| \cdot |K|^3)$ where $|K|$  denotes the number of distinct groups in $F^*$.
\end{restatable}

\begin{proof}
This modification maintains the correctness of the analysis while significantly enhancing the efficiency of solving the underlying flow network. The key improvement arises from the fact that the number of nodes in the network now depends on $|K|$ rather than $|I|$. Consequently, the running time is reduced from $|I| \cdot |C| \cdot |V|^3$ to $|I| \cdot |C|\cdot |K|^3$.
\end{proof}

In practice, the number of categories is typically bounded by a small constant, which also limits the number of distinct groups. As a result, implementing the SCU rule using a compact flow network can significantly reduce the running time, since solving the flow network problem becomes independent of the number of agents.

\begin{example}
\label{example:flow}
Consider a reserve system with six agents $I = \{i_1, i_2, i_3, i_4, i_5, i_6\}$ and three categories $C = \{c_1, c_2, c_3\}$, where $c_2$ is an open category, while $c_1$ and $c_3$ are beneficial categories. Each category has a capacity of 1. The precedence order $\rhd$ dictates that $c_1$ precedes $c_2$, and $c_2$ precedes $c_3$. The priority rankings for each category are as follows:
\[
\begin{aligned}
    c_1: &\quad i_2 \succ i_1 \succ i_3, \\
    c_2: &\quad i_2 \succ i_4 \succ i_6 \succ i_3 \succ i_5 \succ i_1, \\
    c_3: &\quad i_5 \succ i_4 \succ i_6.
\end{aligned}
\]
\end{example}

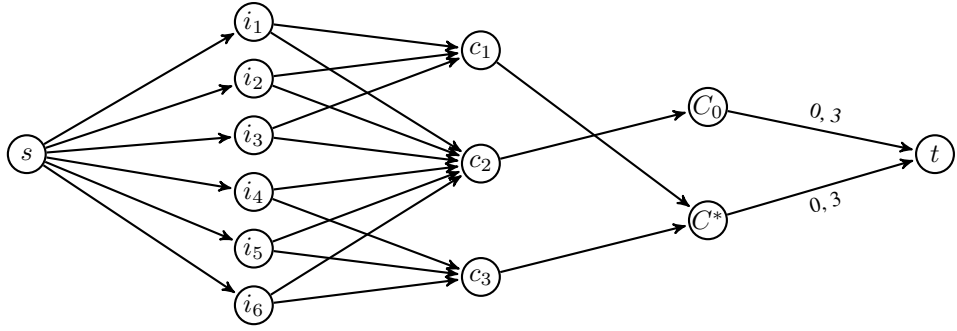
\begin{figure*}[tb]
\begin{center}
\scalebox{1}{ 
\begin{tikzpicture}[-,>=stealth',shorten >=1pt,auto,node distance=1.5cm, 
thick,main node/.style={circle,fill=white!20,draw,minimum size=0.5cm,inner sep=0pt}]

\node [main node] (S) at (-3, -0.25) {$s$}; 
\node [main node] (T) at (9, -0.25) {$t$}; 

\node[main node] at (0, 1.5) (i1)  {$i_{1}$};
\node[main node] at (0, 0.75) (i2)  {$i_2$};
\node[main node] at (0, 0) (i3)  {$i_3$};
\node[main node] at (0, -0.75) (i4)  {$i_4$};
\node[main node] at (0, -1.5) (i5)  {$i_5$};
\node[main node] at (0, -2.25) (i6)  {$i_6$};

\node[main node] at (3, 1.125) (c1)  {$c_{1}$}; 
\node[main node] at (3, -0.375) (c2) {$c_{2}$}; 
\node[main node] at (3, -1.875) (c3) {$c_{3}$}; 

\node[main node] at (6, 0.375) (C0)  {$C_0$}; 
\node[main node] at (6, -1.125) (Cstar)  {$C^*$}; 

\draw[->] (S) -- (i1);  
\draw[->] (S) -- (i2);  
\draw[->] (S) -- (i3);  
\draw[->] (S) -- (i4);  
\draw[->] (S) -- (i5);  
\draw[->] (S) -- (i6);  

\draw[->] (i1) -- (c2);  
\draw[->] (i2) -- (c2);  
\draw[->] (i3) -- (c2);  
\draw[->] (i4) -- (c2);  
\draw[->] (i5) -- (c2);  
\draw[->] (i6) -- (c2);

\draw[->] (i1) -- (c1);  
\draw[->] (i2) -- (c1);  
\draw[->] (i3) -- (c1);  
\draw[->] (i4) -- (c3); 
\draw[->] (i5) -- (c3);  
\draw[->] (i6) -- (c3);

\draw[->] (c1) -- (Cstar);
\draw[->] (c2) -- (C0);
\draw[->] (c3) -- (Cstar);

\draw[->] (C0) -- (T) node[midway, above, rotate=-15, scale=0.8] {0, 3};
\draw[->] (Cstar) -- (T) node[midway, below, rotate=15, scale=0.8] {0, 3};

\end{tikzpicture}
}
\end{center}
\caption{A network flow for Example~\ref{example:flow}, where each edge is associated with a pair $(\ell, u)$, with $\ell$ denoting the lower bound and $u$ the upper bound. Pairs $(0, 1)$ are omitted for brevity.}
\label{fig:network:example}
\end{figure*}

\begin{figure*}[tb]
\begin{center}
\scalebox{1}{ 
\begin{tikzpicture}[-,>=stealth',shorten >=1pt,auto,node distance=1.5cm, 
thick,main node/.style={circle,fill=white!20,draw,minimum size=0.5cm,inner sep=0pt}]

\node [main node] (S) at (-3, 0) {$s$}; 
\node [main node] (T) at (9, 0) {$t$}; 

\node[main node] at (0, 0.5) (g1)  {$k_1$}; %
\node[main node] at (0, -0.5) (g2)  {$k_2$}; %

\node[main node] at (3, 1) (c1)  {$c_{1}$}; 
\node[main node] at (3, 0) (c2) {$c_{2}$}; 
\node[main node] at (3, -1) (c3) {$c_{3}$}; 

\node[main node] at (6, 0.5) (C0)  {$C_0$}; 
\node[main node] at (6, -0.5) (Cstar)  {$C^*$}; 

\draw[->] (S) -- (g1)  node[midway, above, rotate=15, scale=0.8] {0, 3};  
\draw[->] (S) -- (g2) node[midway, below, rotate=-15, scale=0.8] {0, 3};   

\draw[->] (g1) -- (c1) node[midway, above, rotate=15, scale=0.8] {0, 3};  
\draw[->] (g1) -- (c2) node[midway, above, rotate=-15, scale=0.8] {0, 3};
\draw[->] (g2) -- (c2) node[midway, below, rotate=15, scale=0.8] {0, 3};
\draw[->] (g2) -- (c3) node[midway, below, rotate=-15, scale=0.8] {0, 3};

\draw[->] (c1) -- (Cstar);
\draw[->] (c2) -- (C0);
\draw[->] (c3) -- (Cstar);

\draw[->] (C0) -- (T) node[midway, above, rotate=-15, scale=0.8] {0, 3};
\draw[->] (Cstar) -- (T) node[midway, below, rotate=15, scale=0.8] {0, 3};

\end{tikzpicture}
}
\end{center}
\caption{A compact network flow representation for Example~\ref{example:flow}, where agents are partitioned into two groups based on their eligible categories. Specifically, group $k_1$ consists of agents $\{i_1, i_2, i_3\}$, while group $k_2$ consists of agents $\{i_4, i_5, i_6\}$. The lower and upper bounds are adjusted accordingly. Pairs $(0, 1)$ are omitted for brevity.}
\label{fig:network:compact}
\end{figure*}
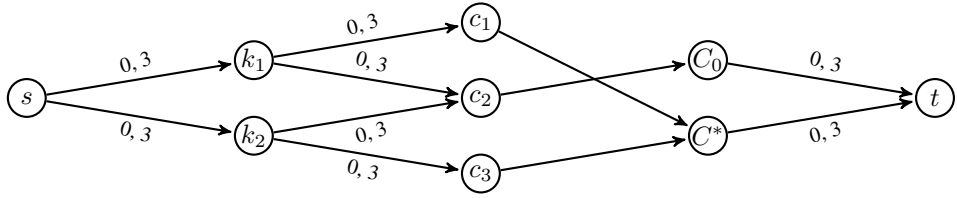

\subsection{Implementation Based on Bipartite Graph}

The challenge in implementing the Sequential Category Update (SCU) rule based on a bipartite graph lies in determining whether there exists a matching in which agent $i$ is assigned to some category $c$ while keeping all agents in $X$ unchanged, ensuring that both maximum properties (maximum cardinality and maximum beneficial assignments) are preserved.

\begin{algorithm}[tb]
\caption{Implementation of the Sequential Category Updating Rule Based on Bipartite Graph}
\label{algorithm:implementation:bipartite}
\SetKwInOut{Input}{Input}
\SetKwInOut{Output}{Output}
\Input{A sequential reserve system $S = (I, C, C^*, q, \unrhd)$}
\Output{A matching $\mu$}
    \SetKw{KwConstruct}{Construct}
    Construct an eligibility graph $G$ and compute a matching $\mu$ satisfying both maximum cardinality and maximum beneficiary assignment\;
    Create a strict order $\rhd$ from $\unrhd$ by breaking ties with the smaller index first\;
    Initialize sets $X \leftarrow \emptyset$ and $Y \leftarrow \emptyset$\;
    \For{each category $c \in C \setminus Y$ in the descending order of $\rhd$}{
        Set $k \leftarrow 1$\;
        \While{$|X \cap \mu^{-1}(c)| \leq q_c$ \textbf{and} $I_c \setminus X \neq \emptyset$}{
            Find the agent $i$ with the $k$-th highest priority from $I_c$\;
            \If{$\mu(i) = c$}{
                Update $X \leftarrow X \cup \{i\}$\;
            }
            \ElseIf{$\mu(i) = \emptyset$}{
                \If{there is an agent $i'$ with the lowest priority in $\mu(c)$ and $i \succ_c i'$}{
                    Update $\mu(i) \leftarrow c$, $\mu(i') \leftarrow \emptyset$, and $X \leftarrow X \cup \{i\}$\;
                }
            }
            \ElseIf{$\mu(i) \in C \setminus \{c\}$}{
                \If{$|\mu(c)| < q_c$}{
                    \If{there exists an alternative path $P$: $c$, $i$, $\mu(i)$, $j$, $\mu(j)$, $\dots$, $\hat{c}$ such that:
                        \begin{itemize}
                            \item it starts with $c$ and ends at $\hat{c}$,
                            \item no agents in $P$ belong to $X$ and no categories in $P$ belong to $Y$,
                            \item both $c$ and $\hat{c}$ are from beneficial or open categories,
                        \end{itemize}}{
                        Update $\mu \leftarrow \mu \oplus P$ and $X \leftarrow X \cup \{i\}$\;
                    }
                }
                \Else{
                    \If{there exists an alternative path $P$: $\hat{c}$, $j$, $\mu(j)$, $\dots$, $c'$, $i'$, $c$, $i$, $\mu(i)$ such that:
                        \begin{itemize}
                            \item it starts with $\hat{c}$ and ends at $\mu(i)$,
                            \item no agents in $P$ belong to $X$ and no categories in $P$ belong to $Y$,
                            \item both $\hat{c}$ and $\mu(i)$ are from beneficial or open categories,
                            \item agent $i'$ has lower priority than $i$ for category $c$,
                        \end{itemize}}{
                        Update $\mu \leftarrow \mu \oplus P$ and $X \leftarrow X \cup \{i\}$\;
                    }
                }
            }
            Set $k \leftarrow k + 1$\;
        }
        Update $Y \leftarrow Y \cup \{c\}$\;
    }
    \Return{The final matching $\mu$}\;
\end{algorithm}

Before introducing the new algorithm, we review some fundamental concepts from graph theory. In a graph $G$, a matching $M$ is a set of edges such that no two edges share a common vertex. An \emph{alternating path} with respect to $M$ is a sequence of edges that begins at an unmatched vertex and alternates between edges in $M$ and edges not in $M$. An \emph{augmenting path} is an alternating path that starts and ends at unmatched vertices. Given a matching $M$ and a path $P$, the operation $M \oplus P$ produces a new matching by adding the edges in $P \setminus M$ and removing the edges in $M \cap P$.

We now present an alternative implementation of the  SCU algorithm using a bipartite graph, as described in Algorithm~\ref{algorithm:implementation:bipartite}.  
First, we compute a matching that satisfies both maximum properties. The algorithm then processes each category in the precedence order defined by $\rhd$ over the set of categories $C$. For each category $c$, it iteratively applies the following checks until either $q_c$ agents from $I_c$ have been added to $X$ or no remaining eligible agents are available.
\begin{description}
    \item[Case 1:] If agent $i$ is already matched to category $c$ in $\mu$, then $i$ is added to $X$, and the current matching $\mu$ remains unchanged.
    \item[Case 2:] If agent $i$ is unmatched, the algorithm searches for the agent $i'$ with the lowest priority in $\mu(c)$ who has a lower priority than $i$ for category $c$. If such an agent exists, $i$ replaces $i'$ at category $c$, updating the matching accordingly, and $i$ is added to $X$.
    \item[Case 3:] If agent $i$ is matched to a category $c' \neq c$, there are two possible situations.
\begin{itemize}
    \item In the first situation, if category $c$ is not full, the algorithm searches for an alternating path $P$: $c$, $i$, $\mu(i)$, $j$, $\mu(j)$, $\dots$, $\hat{c}$. This path must satisfy three conditions: (i) it starts at category $c$ and ends at category $\hat{c}$, (ii) no agents in $P$ belong to $X$ and no categories in $P$ belong to $Y$, and (iii) both $c$ and $\hat{c}$ belong to either the beneficial or open categories. If such a path exists, the matching $\mu$ is updated as $\mu \leftarrow \mu \oplus P$, and agent $i$ is added to $X$.
\item In the second situation, if category $c$ is full, the algorithm searches for an alternative path $P$: $\hat{c}$, $j$, $\mu(j)$, $\dots$, $c'$, $i'$, $c$, $i$, $\mu(i)$, where the following conditions hold: (i) the path starts at $\hat{c}$ and ends at the current match $\mu(i)$, (ii) no agents in $P$ belong to $X$ and no categories in $P$ belong to $Y$, (iii) both $\hat{c}$ and $\mu(i)$ belong to either the beneficial or open categories, and (iv) agent $i'$ has a lower priority than $i$ for category $c$. If such a path exists, the matching $\mu$ is updated as $\mu \leftarrow \mu \oplus P$, and agent $i$ is added to $X$.
\end{itemize}
\end{description}

\begin{restatable}{theorem}{theoremImplementationBipartite}
    \label{theorem:implementation:bipartite}
Given a sequential reserve system $S$ and its corresponding bipartite graph $G$, Algorithm~\ref{algorithm:implementation:bipartite} computes a matching $\mu$ that corresponds to the assignment produced by the SCU rule in time $O(|I| \times |C| \times |E|)$, where $|E|$ denotes the number of edges in $G$.
\end{restatable}

\begin{proof}
We now formally prove that Algorithm~\ref{algorithm:implementation:bipartite} produces the same matching as the SCU rule.

The SCU rule begins with a matching $\mu$ that satisfies both the conditions of maximum cardinality and maximum beneficiary assignment. Let $X$ denote the set of agents whose assignments have been finalized. The algorithm processes each category $c$ in descending order of $\rhd$ and, for each eligible agent $i \in I_c$, iterates over the agents in the order of $\succ_c$. For each agent $i$, the algorithm determines whether $i$ can be matched to category $c$. This determination is made by checking whether an alternative matching $\mu'$ exists that satisfies the following conditions:
\begin{enumerate} 
    \item The assignments of all agents in $X$ remain unchanged.
    \item Agent $i$ is assigned to category $c$.
    \item The matching $\mu'$ preserves the properties of maximum cardinality and maximum beneficiary assignment.
\end{enumerate}

Algorithm~\ref{algorithm:implementation:bipartite} follows a similar procedure. The challenge lies in constructing the alternative matching $\mu'$. If such a matching exists, we consider the following cases:

\begin{description}
    \item[Case 1: Agent $i$ is already matched in $\mu$.] In this case, no modification is necessary, as $\mu$ already satisfies the conditions for matching agent $i$ to category $c$.

    \item[Case 2: Agent $i$ is unmatched in $\mu$.] 
    If an alternative matching $\mu'$ exists, the symmetric difference between $\mu$ and $\mu'$ forms an alternating path $P$: $i, c, i'$, where $i' \in \mu(c)$ (i.e., currently matched to category $c$) has a lower priority than $i$ for category $c$. This implies that $i$ can replace $i'$ in the matching $\mu'$, preserving the conditions.

    \item[Case 3: Agent $i$ is matched to a different category $c' \neq c$.] In this case, we distinguish between two possibilities:

    \begin{itemize}
        \item \textbf{Situation 1:} If category $c$ is not full in $\mu$, then the symmetric difference between $\mu$ and $\mu'$ forms an alternating path $P$: $c, i, \mu(i), j, \mu(j), \dots, \hat{c}$. This path satisfies the following conditions:
        \begin{enumerate}
            \item The path starts at category $c$ and ends at category $\hat{c}$, maintaining maximum cardinality.
            \item No agents in the path belong to $X$, and no categories in the path belong to $Y$, as the assignments of agents in $X$ remain unchanged under both $\mu$ and $\mu'$.
            \item Both categories $c$ and $\hat{c}$ are either beneficial or open categories, ensuring maximum beneficiary assignment.
        \end{enumerate}
        In this case, we update the matching as $\mu \leftarrow \mu \oplus P$, and agent $i$ is added to $X$.

        \item \textbf{Situation 2:} If category $c$ is full in $\mu$, then the symmetric difference between $\mu$ and $\mu'$ forms an alternating path $P$: $\hat{c}, j, \mu(j), \dots, c', i', c, i, \mu(i)$. The following conditions hold for this path:
        \begin{enumerate}
            \item The path starts at category $\hat{c}$ and ends at the current match $\mu(i)$. This happens because agent $i$ replaces some agent $i'$ with a lower priority, who is matched to $c'$. For maximum cardinality, $i'$ is then reassigned to another category, which is a vacancy at $c'$.
            \item No agents in the path belong to $X$, and no categories in the path belong to $Y$.
            \item Both categories $\hat{c}$ and $\mu(i)$ are either beneficial or open categories.
            \item Agent $i'$ has a lower priority than agent $i$ for category $c$.
        \end{enumerate}
        In this case, we update the matching as $\mu \leftarrow \mu \oplus P$, and agent $i$ is added to $X$.
    \end{itemize}
\end{description}

After analyzing all possible scenarios, we conclude that Algorithm~\ref{algorithm:implementation:bipartite} produces the same final matching as the SCU rule.

Next, we analyze the running time. The first step is to compute a matching that achieves both maximum cardinality and maximum beneficiary assignment. By Lemma~\ref{lemma:two_maximum}, we need to compute two maximum matchings using the \textit{Hopcroft-Karp algorithm}. The time complexity for finding a maximum matching in a bipartite graph is $O(|V| \sqrt{|E|})$, where $|V| = |I| + |C|$ is the total number of vertices (agents and categories combined), and $|E|$ is the number of edges.

The second step involves updating the matching. The SCU rule processes each agent $i \in I$ and each category $c \in C$, resulting in $|I| \times |C|$ rounds. In each round, we check for an alternating path, which involves examining edges in the bipartite graph. This step takes $O(|E|)$ time, where $|E|$ is the number of edges in the graph.

Thus, the total time complexity is $O(|V| \sqrt{|E|}) + O(|I| \times |C| \times |E|)$ bounded by $O(|I| \times |C| \times |E|)$.

This completes the proof of Theorem~\ref{theorem:implementation:bipartite}.
\end{proof}

\section{Conclusion}
\label{sec:conclusion}
In this paper, we investigate the health rationing problem, where categories can be processed both simultaneously and sequentially, and agents have dichotomous preferences. Our objective is to achieve the four fundamental properties: eligibility compliance, non-wastefulness, respect for priorities, and maximum cardinality.
The Deferred Acceptance (DA) algorithm is a classical approach that achieves the first three axioms. When agents have strict preferences over categories, rather than dichotomous preferences, the DA algorithm may be a suitable solution.
The Reverse Rejecting (REV) algorithm is the first known algorithm to satisfy all four fundamental axioms. However, it is not as computationally efficient as the MMA algorithm.
The Maximum Matching Adjustment (MMA) algorithm, while straightforward, offers a faster alternative to the REV rule. It is a compelling choice when the primary focus is on the four fundamental properties without additional complexities.
The Sequential Category Updating (SCU) rule is versatile, handling both simultaneous and sequential reserve allocation. It satisfies all the desired axioms and properties considered in this paper. Furthermore, when implemented using flow networks, the SCU rule demonstrates significant computational efficiency, as solving the underlying flow network problem is independent of the number of agents.

\bibliographystyle{plainnat}
\bibliography{reserve}

\end{document}